\documentclass[sigplan,nonacm]{acmart}
\usepackage{amsmath}
\usepackage{array}
\usepackage{booktabs}
\usepackage{graphicx}
\usepackage{multirow}
\usepackage{listings}
\usepackage{tikz}
\usetikzlibrary{arrows.meta,calc,positioning}
\IfFileExists{libertine.sty}{}{\usepackage{lmodern}}

\lstdefinestyle{cuda}{
  language=C++,
  morekeywords={__global__,__device__,__shared__,__restrict__,__syncthreads,
    __launch_bounds__,blockIdx,threadIdx,blockDim,gridDim,cudaMalloc,cudaFree,
    cudaMemcpy,cudaMemset,uint32_t,int32_t,size_t,float4,int4},
  basicstyle=\ttfamily\scriptsize,
  keywordstyle=\color[rgb]{0.13,0.29,0.53}\bfseries,
  commentstyle=\color[rgb]{0.35,0.45,0.35}\itshape,
  stringstyle=\color[rgb]{0.55,0.25,0.10},
  numbers=left, numberstyle=\tiny\color{gray}, numbersep=4pt,
  breaklines=true, breakatwhitespace=false, columns=fullflexible,
  showstringspaces=false, tabsize=2, frame=none, xleftmargin=12pt,
}

\setcopyright{none}
\renewcommand\footnotetextcopyrightpermission[1]{}
\acmConference[Anonymous Submission]{Anonymous Submission}{2026}{}
\acmYear{2026}

\begin{document}

%\title[]{SparseDitto: An Agentic Sparse Compilation Framework through Architecture-Aware Synthesis on GPUs}
\title[SparseDitto: An Agentic Sparse Compilation Framework through Architecture-Aware Synthesis on GPUs]{ \texorpdfstring{\raisebox{-0.35\height}{\includegraphics[height=35pt]{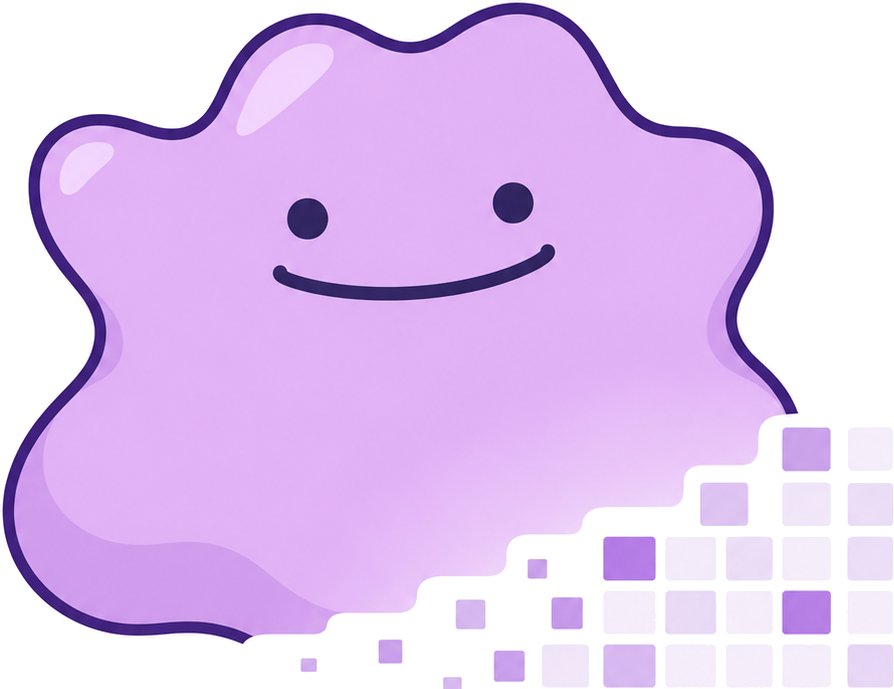}}\hspace{0.3em}}{} SparseDitto: An Agentic Sparse Compilation Framework through Architecture-Aware Synthesis on GPUs}

% \author{Anonymous Author(s)}
% \affiliation{%
%   \institution{Anonymous Institution}
%   \city{Anonymous City}
%   \country{Anonymous}}
% \renewcommand{\shortauthors}{Anonymous Author(s)}

\author{Shiyang Li}
\email{li004074@umn.edu}
\affiliation{%
  \institution{University of Minnesota }
  \city{Minneapolis}
  \state{MN}
  \country{USA}}
\author{Guangyan Sun}
\email{sun01158@umn.edu}
\affiliation{%
  \institution{University of Minnesota }
  \city{Minneapolis}
  \state{MN}
  \country{USA}}
\author{Jinwei Tang}
\email{tang0940@umn.edu}
\affiliation{%
  \institution{University of Minnesota }
  \city{Minneapolis}
  \state{MN}
  \country{USA}}
\author{Yanzhi Wang}
\email{yanz.wang@northeastern.edu}
\affiliation{%
  \institution{Northeastern University}
  \city{Boston}
  \state{MA}
  \country{USA}}
\author{Mingyi Hong}
\email{mhong@umn.edu}
\affiliation{%
  \institution{University of Minnesota }
  \city{Minneapolis}
  \state{MN}
  \country{USA}}
\author{Caiwen Ding}
\email{dingc@umn.edu}
\affiliation{%
  \institution{University of Minnesota }
  \city{Minneapolis}
  \state{MN}
  \country{USA}}
\renewcommand{\shortauthors}{Li et al.}
\begin{abstract}
Sparse matrix computation performance on GPU depends on how representation and execution schedule match the input structure and target hardware. No single implementation consistently dominates across sparsity patterns, operators, and hardwares.  Existing sparse compilers and specialized systems cannot cover all of them simultaneously.

We present SparseDitto, an agentic sparse compilation framework for sparse matrix computation on GPUs. It jointly synthesizes representation, execution schedule, and hardware mapping in a unified compilation plan. Structural analysis and a learned template-ranking prior guide architecture-aware synthesis. LLM-guided lowering realizes each plan as CUDA code, while target-GPU profiling drives plan refinement. SparseDitto covers multiple operators, e.g., SpMV, SpMM, and SpGEMM, and various representations within one framework. It can also automatically adapt to different hardwares. Across various SuiteSparse matrices, SparseDitto achieves geometric-mean speedups over cuSPARSE of $2.68\times$ on an NVIDIA RTX PRO 6000 and $2.79\times$ on an NVIDIA H200 (up to 146.61$\times$). Its generated SpMM kernels accelerate full-batch GCN training by up to $3.39\times$.

\end{abstract}

\keywords{sparse matrix computation, GPU kernel synthesis, sparse compilation framework }
% large language models}

\maketitle

\raggedbottom
\setcounter{topnumber}{1}

\section{Introduction}
\label{sec:intro}

Sparse matrix operations are core building blocks in scientific computing, graph analytics, and machine learning~\cite{VAZQUEZ2010146,TC-GNN,taco,csr5,maxkgnn}. Their performance on GPUs can change by orders of magnitude even when the matrix, operator, library, and GPU remain unchanged. For one matrix with 355\,K nonzeros, cuSPARSE completes SpMM in 37\,$\mu$s using CSR but requires 13\,ms using Blocked-ELL. Blocked-ELL regularizes sparse computation by storing each block row with a common number of blocks, which can expose efficient blocked execution. On this irregular matrix, however, the required padding makes much of the transferred data useless and removes that advantage. The $350\times$ performance difference is caused only by the representation and its associated kernel. This example exposes a basic challenge for sparse matrix computing on GPU. A high-performance sparse implementation needs to match its representation and execution strategy to the structure of the input.
% no detail, the transition between first 2 para. More analysis about this example

\begin{figure}[t]
    \centering
    \includegraphics[width=0.9\linewidth,trim=0 12bp 0 0,clip]{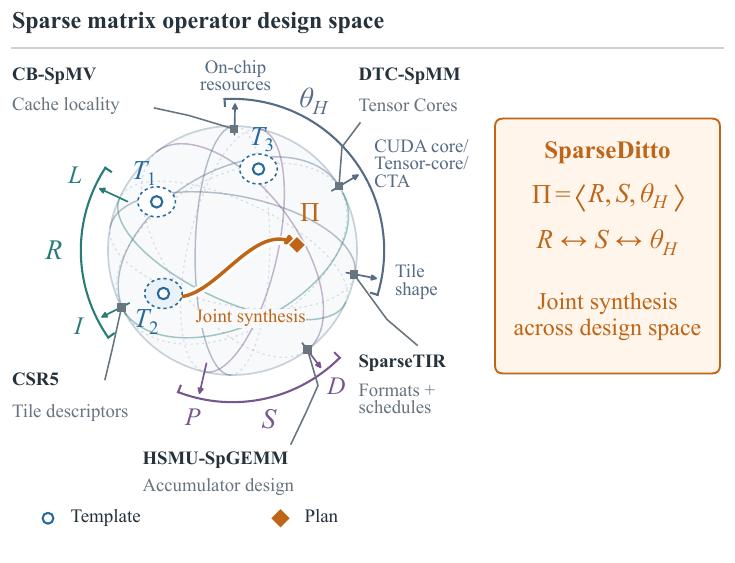}
    \par\smallskip
    (a) Design space of sparse matrix operators\par
    \smallskip
    \includegraphics[width=\linewidth]{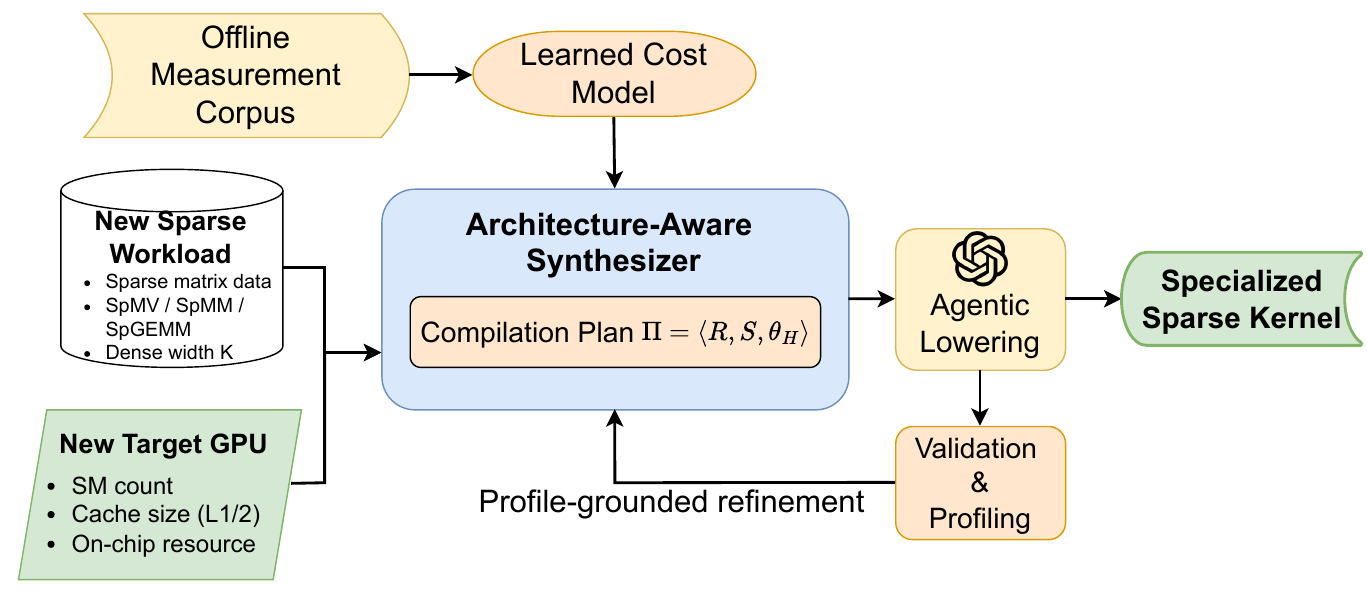}
    \par\smallskip
    (b) SparseDitto workflow\par
    \caption{SparseDitto's design space and compilation workflow. Prior-work annotations in (a) highlight representative mechanisms.}
    \label{fig:intro_overview}
    \Description{Two vertically stacked panels. The first shows the sparse kernel design space and SparseDitto's joint synthesis across representation, execution schedule, and hardware mapping. The second shows its compilation workflow, from a sparse workload and target GPU to a specialized kernel, with profile-grounded refinement.}
\end{figure}

%Across sparse matrix-vector multiplication (SpMV), sparse matrix-dense matrix multiplication (SpMM), and sparse matrix-matrix multiplication (SpGEMM), proper optimization decisions highly depend on sparsity pattern. 
%Row-length distributions affect load balance~\cite{cong2025cb-spmv}, while column indices affect memory locality~\cite{MatRaptor}. Block occupancy determines whether blocked execution is worthwhile~\cite{TC-GNN}. For SpGEMM, the distribution of intermediate products further affects sparse accumulation~\cite{wu2025hsmu,TileSpGEMM,cluster-wise-spgemm}. 
Further fundamental challenge is that sparse GPU performance depends on three coupled implementation choices (Figure~\ref{fig:intro_overview}a). The sparse representation determines the layout of values and metadata. The execution schedule determines workload partitioning, thread mapping, and data movement. Hardware mapping further binds these choices to tile sizes, launch configurations, and on-chip resources on the target GPU.

Current state-of-the-art (SOTA) approaches follow three main directions. First, specialized sparse libraries and kernels encode expert-designed optimizations for particular operators, sparsity patterns, or hardware features~\cite{cong2025cb-spmv,dtcspmm,wu2025hsmu}. Second, sparse compilers and design-space search systems expose broader choices of representations and schedules~\cite{taco,sparsetir,AlphaSparse}. Third, using LLM-based CUDA programming agents to generate and optimize GPU kernels through iterative coding, compilation, profiling, and refinement~\cite{qimeng,kernelbench,stitchcuda}. These directions provide optimization mechanisms, but they partially addressed the challenges. The key missing capability is to coordinate sparse-specific representation and scheduling decisions with flexible code construction and target-specific hardware adaptation.

%State-of-the-art (SOTA) sparse libraries or sparse compilers are typically designed for one sparse kernel with only one specific design choice. 
We study three representative operator-specialized systems and one compiler-based system: CB-SpMV~\cite{cong2025cb-spmv}, DTC-SpMM~\cite{dtcspmm}, HSMU-SpGEMM~\cite{wu2025hsmu}, and SparseTIR~\cite{sparsetir}. We obtain two key observations. \textbf{First, sparse optimization is highly asymmetric.} A compatible representation and schedule can provide meaningful gains, while a poor match can cause orders-of-magnitude slowdowns. The best alternative format improves SpMV by up to $1.47\times$ and SpMM by up to $2.8\times$, while a poor pairing loses up to $350\times$. \textbf{Second, no fixed implementation consistently dominates across sparsity patterns, operators, and GPUs.} CB-SpMV works well on cache-friendly blocked inputs. DTC-SpMM benefits matrices that keep enough useful work inside compressed Tensor-Core blocks. HSMU-SpGEMM performs well when its hash accumulator matches the intermediate-product distribution. Their relative performance also changes with the GPU architecture and CUDA runtime version.

% A fixed implementation therefore embeds assumptions about both the sparse structure and the target platform.
The three SOTA directions leave complementary gaps. Specialized kernels cover only the sparsity patterns, operators, and hardware assumptions for which they were designed. Sparse compilers and design-space search systems broaden the optimization space, but their coverage is still bounded by developer-provided static rules. Search can also be expensive. AlphaSparse~\cite{AlphaSparse}, for example, can spend hours constructing a specialized SpMV implementation through offline search. General CUDA programming agents offer more flexible code construction, but they lack sparse-specific guidance for choosing representations and execution schedules. A functionally correct kernel can therefore remain inefficient when these high-level decisions do not match the input structure. 
A general solution needs both structured sparse optimization and the ability to adapt to the target GPU.
% Existing sparse compilers and design space search systems provide useful mechanisms for expressing or exploring sparse optimizations~\cite{taco,sparsetir,AlphaSparse}, but they still need a way to construct an effective compilation plan for each case. Such a plan must choose the sparse representation, the execution schedule, and the hardware mapping together. It must also adapt these choices when the operator or target GPU changes. This motivates automatic synthesis of the compilation plan itself.

Our target is to end the game of designing a separate sparse optimization for every limited use case and hardware platform. We seek a framework that can combine existing sparse optimization strategies, use each where it fits, and adapt to new GPU platforms with minimal additional human effort. This goal leads to three requirements. The framework must jointly reason about representation, execution schedule, and hardware mapping. It must use structural analysis to guide these choices. It must also incorporate target-GPU measurements when static analysis is insufficient.

In this paper, we present \textbf{SparseDitto}, the first agentic sparse compilation framework that automatically specializes GPU kernels for a given sparse matrix, operator, and target GPU. As shown in Figure~\ref{fig:intro_overview}b, SparseDitto treats sparse kernel construction as a compilation problem. Architecture-aware synthesis builds a unified plan for representation, execution schedule, and hardware mapping. Agentic lowering realizes the plan as a specialized CUDA implementation for repeated execution. Subsequent calls reuse it for the same sparsity structure and operator configuration on the target GPU.
%Structural analysis first extracts properties of the sparse input and estimates representation and operator-specific costs. An interpretable learned cost model ranks established optimization templates using an offline measurement corpus. The synthesis stage then combines this prior with static workload bounds and target-GPU constraints to construct candidate compilation plans. An LLM-guided lowering stage turns each plan into CUDA code. Compilation, correctness validation, and target-GPU profiling provide feedback to refine the plan. 

In summary, this paper makes the following contributions:
\begin{itemize}
  \item \textbf{A unified sparse kernel compilation framework on GPUs.}
  We formulate sparse GPU kernel compilation as the joint choice of data representation, execution schedule, and target-specific hardware mapping. The formulation captures the major implementation decisions shared by SpMV, SpMM, and SpGEMM within one framework.

  \item \textbf{Structure-guided and architecture-aware synthesis.}
  A structural analysis computes 36 features that describe the sparsity pattern, representation overhead, and operator-dependent intermediate work. An interpretable additive energy model serves as a learned cost model and ranks established optimization templates from an offline measurement corpus. An architecture-aware synthesis stage combines this ranking with static workload bounds and target-GPU constraints to construct candidate compilation plans.

  \item \textbf{LLM-guided lowering with profile-guided optimization.}
  SparseDitto lowers each compilation plan into CUDA using an LLM-guided coding stage and a set of supported sparse optimization mechanisms. Compilation and numerical validation check each candidate. Runtime measurements and profiling on the target GPU guide later changes to the representation, schedule, and hardware mapping. The fastest valid implementation is retained for reuse.
\end{itemize}

We evaluate SparseDitto on $60$ SuiteSparse matrices, covering SpMV, SpMM at four dense widths, and SpGEMM. SparseDitto reaches a geometric-mean speedup of $2.68\times$ over cuSPARSE on an RTX PRO 6000, with a maximum of $146.61\times$. On an H200, it reaches $2.79\times$, with a maximum of $78.5\times$. The specialized SpMM kernels also accelerate full-batch GCN training by up to $3.39\times$.

\section{Background and Motivation}
\label{sec:background}

%GPU sparse matrix operators combine three coupled implementation decisions: data representation, execution schedule, and hardware mapping. These decisions determine memory traffic, memory footprint, available parallelism, synchronization cost, and hardware utilization.
%The representation specifies the value layout and sparse metadata. The schedule specifies workload partitioning, thread mapping, and dataflow. Hardware mapping selects tile sizes, launch configurations, and on-chip resource use for the target GPU. 

%Consider ELL and SELL-C-$\sigma$. ELL stores all rows in one padded layout. SELL divides rows into slices and may reorder them within a window~\cite{sellcsigma}. The latter adds slice metadata but can reduce padding for irregular row lengths. Its slice height also affects thread mapping on the GPU. Conversely, two kernels can retain CSR and use different schedules. A row-based kernel assigns rows to warps, whereas a merge-based kernel partitions the nonzero stream and performs segmented reductions~\cite{mergespmv}. A format name therefore captures only part of a sparse implementation.

% This section first explains how the sparse operator, sparsity pattern, and target hardware shape the design space (\S\ref{sec:bg-ops}). We then quantify the performance cliffs from mismatched choices (\S\ref{sec:bg-formats}). Finally, we show the main limitation of existing works and how it motivates SparseDitto (\S\ref{sec:bg-strategies}).

\begin{figure*}[ht]
  \centering
  \includegraphics[width=\textwidth]{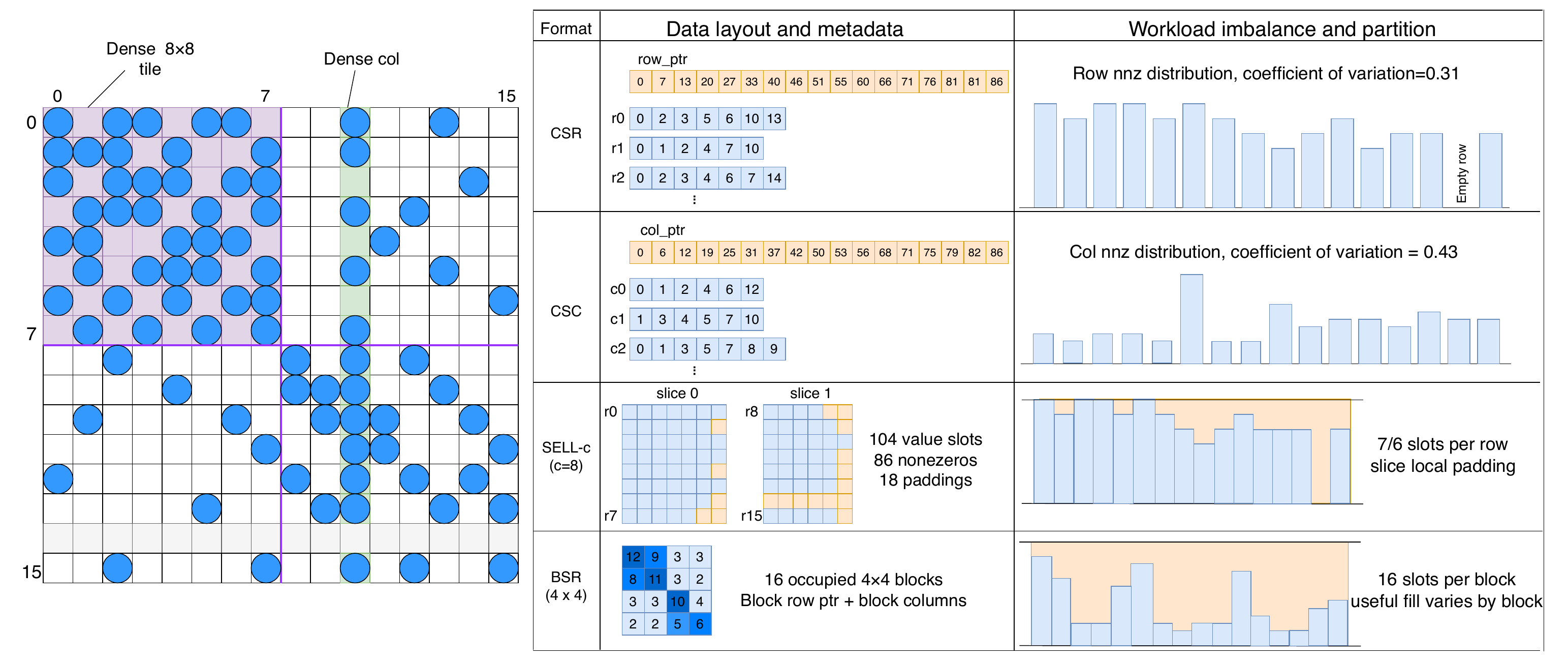}
  \caption{One $16\times16$ matrix exposes intrinsic pattern cues and format-induced costs. The matrix contains an irregular dense tile, a high-degree column, an empty row, and uneven row lengths. CSR and CSC traverse compressed rows and columns, respectively. SELL-C with $C=8$ introduces slice-local padding, while $4\times4$ BSR introduces block fill. The right column shows the resulting work granularity and
  imbalance for rows, columns, slice slots, and blocks.}
  \label{fig:format-illustration}
\end{figure*}

\subsection{Sparse Matrix Operators on GPUs}
\label{sec:bg-ops}

The sparsity pattern determines the distribution of nonzeros, the memory a kernel traverses, and the workload assigned to computing units. It also controls locality once the schedule is mapped to the GPU. The operator determines the work produced by each visited nonzero. As a result, the critical design choice changes across SpMV, SpMM, and SpGEMM.

SpMV computes $y=Ax$ and performs only two floating-point operations
per nonzero. Its performance often depends on sparse metadata traffic and the locality of gathers from $x$. Load imbalance among parallel units is another major cost~\cite{coo,csr5,mergespmv}. Row-based CSR avoids output reductions but inherits the row-length distribution. ELL and SELL use padding to regularize sparse traversal. 
% while BSR reduces index traffic when occupied blocks have sufficient fill.

SpMM computes $C=AB$ with a dense matrix $B$ of width $K$. Each nonzero of $A$ drives $K$ multiply--adds, which creates another pattern dimension and more reuse of $B$. The sparse representation needs to be coordinated with the mapping of rows or blocks to the GPU. The $K$-tile and output-accumulator placement also affect the dataflow. Dense sparse tiles can enable Tensor Cores when useful arithmetic outweighs compaction and fill-in costs~\cite{TC-GNN,dtcspmm}.

SpGEMM further discovers a sparse output. It computes $C=AB$ for two sparse matrices inputs $A$ and $B$. In a row-wise algorithm, row $i$ generates
\begin{equation}
  u_B(i)=\sum_{k\in A[i,:]} nnz(B[k,:]),
  \label{eq:bg-spgemm-ub}
\end{equation}
The distribution of $u_B(i)$ affects row
partitioning and temporary storage. It also determines the suitability of dense, hash, merge, or hybrid accumulators~\cite{MatRaptor,nsparse,wu2025hsmu}. Design space now extends beyond input traversal because the kernel must construct output indices and intermediate state.

Thus, the same pattern can produce different metadata streams, processing units, and padding or fill costs. Figure~\ref{fig:format-illustration} shows these effects on a common matrix. They are coupled to the parallel decomposition selected by the GPU kernel. 

These interactions motivate two questions. First, with the operator and GPU fixed, how strongly does the profitable implementation depend on the sparsity pattern? Second, how much of this pattern-dependent optimization space is covered by existing specialized systems and sparse compilers?

%rather than determined by the data format alone.

% We answer the first question in \S\ref{sec:bg-formats} and the second in \S\ref{sec:bg-strategies}.

\subsection{Sensitivity to Sparsity Patterns}
\label{sec:bg-formats}
\begin{figure*}[ht]
  \centering
  \includegraphics[width=\textwidth,trim=0 6bp 0 2bp,clip]{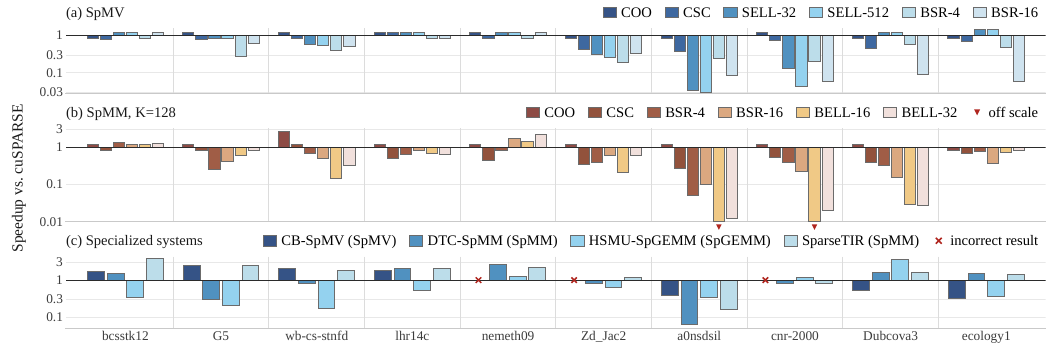}
  \caption{Sparse kernel performance sensitivity on ten SuiteSparse
  matrices~\cite{suitesparse}. Top: cuSPARSE implementations for
  different sparse representations in SpMV and SpMM at $K{=}128$.
  Bottom: three specialized systems and one sparse compiler. The best design changes
  with the pattern, while mismatched choices create severe slowdowns.
  All measurements use an NVIDIA RTX PRO 6000 Blackwell GPU with CUDA 13.0
  and report the mean of ten timed iterations after warmup.}
  \label{fig:background-characterization}
\end{figure*}

We first hold the library and platform fixed to cuSPARSE~\cite{cusparse}, then vary the representation and SpMM width. The top panel of Figure~\ref{fig:background-characterization} uses format names for brevity, but every bar measures a complete implementation. 
%The measurement includes its metadata traversal and parallel mapping. It also includes the library-selected kernel configuration.

The observed spread follows the assumptions of each implementation.
SELL-C wins by up to $1.47\times$ on a stencil with only 0.08\% slice padding. It loses $33\times$ when uneven row lengths inflate storage by $30\times$. BSR wins $1.25\times$ when occupied $16{\times}16$ tiles are 87\% dense. It falls to $0.04\times$ on power-law graphs because the occupied blocks contain mostly zeros. COO pays for explicit row metadata and reductions on regular inputs. Under row imbalance, its nonzero-level decomposition makes it the most robust SpMM choice. COO wins 21 of 30 configurations and reaches $2.8\times$. Across the full sweep, the best alternative gains at most $2.8\times$ over CSR. A mismatched design loses up to $350\times$.

The operator context changes the same tradeoff. At $K{=}32$, block-dense matrices can favor CUDA-Core blocked kernels. At $K{=}128$, Tensor-Core layouts replace them on some inputs. Blocked-ELL accelerates dense-tile arithmetic but pads every block row to a common width. This padding can increase transferred bytes by two orders of magnitude. For SpGEMM, the corresponding choices appear in row bucketing and phase structure. The accumulator dataflow also depends on the intermediate-work distribution ~\cite{TileSpGEMM,nsparse,speck,wu2025hsmu}.

\textbf{Observation 1.} The profitability of a sparse optimization depends strongly on the input structure. Representation and execution schedule must therefore be coordinated rather than chosen independently. A poor match can be far more costly than the gain from a good one.

\subsection{Limitations of Existing Works}
\label{sec:bg-strategies}
We next evaluate three recent operator-specialized systems and one sparse compiler on the same NVIDIA RTX PRO 6000 GPU platform. CB-SpMV~\cite{cong2025cb-spmv} combines cache-friendly blocks with adaptive per-block formats. DTC-SpMM~\cite{dtcspmm} combines compressed block metadata with Tensor-Core execution. HSMU-SpGEMM~\cite{wu2025hsmu} uses a shared-memory hash accumulator for SpGEMM. SparseTIR~\cite{sparsetir} supports dynamic format decomposition and scheduling for SpMM. 
%CB-SpMV fails numerical validation on three matrices, so we exclude those timings.

Figure~\ref{fig:background-characterization} shows how they perform. CB-SpMV achieves $1.7$--$2.5\times$ while its blocked working set remains cache-friendly. It loses on every matrix above 3\,M nonzeros and falls to $0.32\times$. DTC-SpMM reaches $3.1\times$ on block-structured inputs. It also beats cuSPARSE's padded Tensor-Core format by $26$--$106\times$ on irregular matrices. However, under extreme row imbalance DTC-SpMM falls to $0.04\times$. HSMU-SpGEMM surpasses cuSPARSE on three and loses on seven of ten matrices. None of the three systems dominates the full set.

The target platform changes the useful region further. These systems were developed for earlier GPU and CUDA generations. On Blackwell and CUDA 13.0, their fixed resource choices meet different hardware conditions, and they adapt poorly to the new platform. DTC-SpMM, for example, reports more than $1.5\times$ over cuSPARSE on an RTX 4090~\cite{dtcspmm}. Running the released artifact on our Blackwell GPU with CUDA 13.0, we measure it below cuSPARSE at every dense width, from $0.73\times$ at $K{=}8$ down to $0.58\times$ at $K{=}256$. 
%A sparse strategy must therefore be configured and validated on its target GPU.

Sparse tensor compilers expose a broader space through dynamic format decomposition. They also provide scheduling primitives. SparseTIR~\cite{sparsetir} beats cuSPARSE on 25 of 30 SpMM configurations. Its geometric-mean speedup is $1.40\times$ and its maximum is $5.0\times$. SparseTIR nevertheless falls to
$0.13$--$0.20\times$ on the matrix behind the $350\times$ format cliff because long rows serialize execution. SparseTIR can express a bucketed hybrid layout, but a developer supplies the format rewrite and schedule~\cite{sparsetir}. Its coverage therefore depends on the implementations encoded in its rule set.
It also does not support SpMV or SpGEMM, and it covers only a limited set of widths ($K=32$, $128$, $256$) for SpMM.

\textbf{Observation 2.} No fixed implementation or developer-defined optimization space consistently covers different sparsity patterns, operators, and GPU architectures. The useful optimization region changes with both the input structure and the target platform.

Together, Observations 1 and 2 show that sparse kernel optimization is both pattern dependent and architecture dependent. This motivates treating sparse kernel construction as a per-workload compilation problem. Representation, execution schedule, and hardware mapping should be constructed together for the input structure and target GPU.

\section{SparseDitto Design}
\label{sec:design}

SparseDitto treats sparse GPU kernel construction as a per-workload compilation problem. A compilation task is specified by a sparse input $X$, an operator $o$, and a target GPU $H$. Given $\langle X,o,H\rangle$, SparseDitto constructs a specialized CUDA implementation together with the sparse data structures required by that implementation.

SparseDitto represents the main compilation decisions with a \emph{compilation plan}
\begin{equation}
    \Pi = \langle R,S,\theta_H\rangle,\qquad
    R=\langle L,I\rangle,\qquad
    S=\langle P,D\rangle .
    \label{eq:compilation-plan}
\end{equation}
The representation $R$ consists of a data layout $L$ and sparse metadata $I$. The layout specifies how sparse values are organized and stored, while the metadata supports traversal of the resulting representation. The execution schedule $S$ consists of a parallel decomposition $P$ and a dataflow $D$. The decomposition maps work to Cooperative Thread Arrays (CTAs), warps, and threads, while the dataflow determines operand movement and accumulation. Finally, $\theta_H$ contains target-specific choices such as tile sizes, launch configuration, and on-chip resource allocation. These decisions are coupled. A change in any of them can impact others.
%representation can alter the available parallel decomposition and dataflow. It can also change the hardware configuration that is effective on the target GPU. SparseDitto therefore constructs $R$, $S$, and $\theta_H$ together rather than optimizing them independently.

\begin{figure*}[t]
  \centering
  \includegraphics[width=\textwidth]{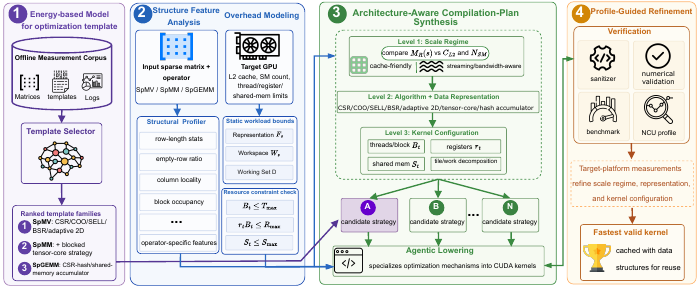}
  \caption{Overview of SparseDitto. Sparse pattern analysis and a learned cost model identify promising optimization templates. Architecture-aware synthesis constructs compilation plans that coordinate representation, execution schedule, and hardware mapping. Agentic lowering and profiling realizes these plans as CUDA code.}
  \label{fig:design-overview}
\end{figure*}

Figure~\ref{fig:design-overview} shows the overall compilation process. SparseDitto first analyzes the sparse input and estimates the overhead. A energy-based model then ranks reusable optimization templates. The architecture-aware synthesis stage combines this ranking with the input analysis and target-GPU constraints to construct candidate compilation plans. Finally, an agentic lowering stage realizes each plan as CUDA. Sanitizer, numerical validation, and target-GPU profiling provide feedback for subsequent plan refinement.

\subsection{Structural Features and Overhead Modeling}
\label{sec:design-profile}

SparseDitto performs a lightweight sparse analysis before compilation-plan synthesis. The analysis is guided by known sparse optimization mechanisms. For each mechanism, we identify the structural conditions that make it effective and estimate the overhead it can incurred, e.g., extra storage. The analysis produces three groups of continuous features:
%The resulting features serve two purposes. 
%They provide input to the energey-based model in \S\ref{sec:design-selector} and quantitative guidance for architecture-aware synthesis in \S\ref{sec:design-planning}.

\begin{equation}
    \Phi(X,o) =
    \left\langle
    \Phi_{\mathrm{intr}},
    \Phi_{\mathrm{repr}},
    \Phi_{\mathrm{op}}
    \right\rangle .
    \label{eq:structural-evidence}
\end{equation}
$\Phi_{\mathrm{intr}}$ describes the input sparsity pattern. $\Phi_{\mathrm{repr}}$ estimates the overhead introduced by candidate sparse representations. $\Phi_{\mathrm{op}}$ captures operator-dependent work that is not visible from the input structure alone. Together, these groups contain 36 continuous features.

\noindent\textbf{Intrinsic pattern features.}~Let $r_i=\operatorname{nnz}(X[i,:])$ denote the number of nonzeros in row $i$. Matrix size and density characterize the overall sparse working set. SparseDitto records the mean, maximum, and coefficient of variation (CV) of $r_i$. It also records row-length quartiles and the fraction of empty rows. These statistics expose workload imbalance relevant to CSR5, merge-based decomposition, and SELL-$C$-$\sigma$~\cite{csr5,mergespmv,sellcsigma}.

SparseDitto also measures structure along the column dimension. Column-degree statistics capture reuse and asymmetry between rows and columns. Row span and normalized row span characterize access locality. We additionally record the mean gap between sorted column indices and the fraction of diagonal entries. To capture block structure, SparseDitto measures the coverage of occupied $16\times16$ tiles and the fraction of nonzeros contained in them. This group contains 18 continuous features. SparseDitto also assigns the row-length distribution to one of five shape classes: uniform, moderate, skewed, power-law, or bimodal. 
%The shape class is used as a categorical input by the learned cost model.

\noindent\textbf{Representation-induced overhead features.}~The representation $R=\langle L,I\rangle$ changes both the working set size and the metadata traversed by a kernel. SparseDitto computes value-slot expansion for padded and blocked layouts:
\begin{align}
 \rho_{\mathrm{ELL}} &={m\max_i r_i}/{\operatorname{nnz}(X)}, \notag\\
 \rho_{\mathrm{SELL}}(c) &=
   {\textstyle\sum_{s}|s|\max_{i\in s}r_i}/{\operatorname{nnz}(X)}, \notag\\
 \rho_{\mathrm{BSR}}(b) &={b^2 \operatorname{nnzb}_b}/{\operatorname{nnz}(X)},
 \label{eq:format-expansion}
\end{align}
where $s$ denotes a row slice and $\operatorname{nnzb}_b$ counts occupied $b{\times}b$ blocks. These factors estimate padded computation and value traffic. The storage estimate also accounts for slice offsets and block-column indices. Row permutations are included when required. SparseDitto evaluates ELL and three SELL slice heights. It also evaluates BSR with $b\in\{4,8,16\}$. These choices produce eight continuous representation features. Their values are recomputed when $L$, $I$, or the format parameters change during plan refinement.

\noindent\textbf{Operator-induced features.}~Input statistics alone do not expose the intermediate iteration space of SpGEMM. From Equation~\ref{eq:bg-spgemm-ub}, SparseDitto computes the total, maximum, p95, and CV of $u_B(i)$ for the requested product. It also computes the total and CV for $B=X^T$. The accumulator-fit statistic
\begin{equation}
 F_h=\frac{1}{m}\sum_i \mathbf{1}[u_B(i)\le h]
 \label{eq:accumulator-fit}
\end{equation}
reports the fraction of rows below capacities $h=512$, 4K, and 32K. These thresholds distinguish accumulator regimes that a mean would hide. SparseDitto also computes
\begin{equation}
 \gamma_B={\textstyle\sum_i u_B(i)}/{\operatorname{nnz}(C)},
 \label{eq:compression-ratio}
\end{equation}
which measures the intermediate products merged into each output nonzero. The estimator uses stratified sampling inspired by dataflow analysis in~\cite{MatRaptor}. The ten features in this group capture intermediate work, imbalance, accumulator demand, and output compression.

\noindent\textbf{Hardware and task context.}~Target-GPU properties are kept separate from the 36 sparse features. SparseDitto records the number of SMs and physical cache capacities in bytes. It also records per-CTA hardware limits on threads, registers, and shared-memory size. The operator and the dense width $K$ for SpMM specify the task context. These values are used by the learned cost model when applicable and by the architecture-aware synthesis stage to construct and check compilation plans.

%The feature set is derived from established sparse optimization mechanisms~\cite{MatRaptor,sextans,maxkgnn,cong2025cb-spmv,dtcspmm,TC-GNN,sparsetir,AlphaSparse,zhao2018bridging,sellcsigma,csr,csr5,cusparse,coo,nsparse}. SparseDitto computes these features using bounded scans, occupied-block counts, and budgeted sampling. 

\subsection{Energy-Based Model for Optimization Templates}
\label{sec:design-selector}

The learned cost model provides a prior over reusable optimization templates. We adopt an additive parameterization so that each ranking can be decomposed into feature-level contributions and passed to compilation-plan synthesis. It uses measured experience to rank template set $V_o$, that has a large effect on representation overhead, workload decomposition, or dataflow~\cite{sedaghati2015automatic,zhao2018bridging,dtcspmm,wu2025hsmu,sparsetir,cong2025cb-spmv,cusparse}. 
%The template set $V_o$ covers coarse implementation families, while the synthesis stage later determines the remaining representation, schedule, and hardware-specific parameters.

For SpMV, $V_o$ contains CSR, COO, SELL with slice heights $\{16,32,512\}$, BSR with block sizes $\{4,8,16\}$, and adaptive 2D blocking. The SpMM set contains CSR, COO, the same three BSR variants, Blocked-ELL-$\{16,32\}$, and the DTC Tensor-Core design~\cite{dtcspmm}. The three SpGEMM templates are CSR-hash, shared-memory tiled hash, and hybrid accumulation. These templates draw on established GPU kernels~\cite{sellcsigma,MatRaptor}.

The learned cost model is a single additive energy model~\cite{lecun2006tutorial, hastie1990gam}. Its input $x$ concatenates the 36 sparse features, a one-hot shape class, an operator indicator, and the tile width $\log_2 K$ for SpMM. For each scalar coordinate $x_j$, a nonlinear basis shared across operators produces
\begin{equation}
  h_j(x_j)=\tanh(x_j w_{1j}+b_{1j})\in\mathbb{R}^{16}.
  \label{eq:energy-basis}
\end{equation}
Each operator $o$ then has its own readout for every template $s\in V_o$:
\begin{equation}
 E_o(s\mid x)=\beta_{o,s}+
   \sum_j h_j(x_j)^\mathsf{T} w_{2,o,j,s}.
 \label{eq:additive-energy}
\end{equation}
Lower energy indicates a stronger recommendation. Sharing the basis across operators lets a single feature response be estimated from all three tasks, while the operator-specific readouts capture the different effect of that feature on SpMV, SpMM, and SpGEMM. The nonlinear basis can express threshold behavior, such as a sharp penalty once SELL padding exceeds a level, without a hand-designed cutoff.

We train the energy directly from measured runtimes. For a training example, let $M\subseteq V_o$ be the templates actually benchmarked and $t_s=\log_2(\text{baseline runtime}/\text{runtime of }s)$ their measured log-speedups. The objective combines a listwise likelihood~\cite{cao2007listnet} with a gap-weighted pairwise margin~\cite{burges2005ranknet}:
\begin{equation}
\small
\begin{aligned}
  \mathcal{L}_o=\sum_{(x,\mathbf{t})}\bigg[
    &-\sum_{s\in M} q_s \log p_o(s\mid x)\\
    &+\lambda_o\!\!\sum_{\substack{s,s'\in M\\ t_s>t_{s'}}}\!\!
       (t_s-t_{s'})\,
       \operatorname{softplus}\!\big(E_o(s\mid x)-E_o(s'\mid x)\big)
  \bigg]
\end{aligned}
\label{eq:selector-loss}
\end{equation}
where $p_o(s\mid x)=\mathrm{softmax}_{s\in M}\!\big(-E_o(s\mid x)\big)$ and $q_s=\mathrm{softmax}_{s\in M}\!\big(t_s/T\big)$ are soft targets from the measured speedups. The listwise term keeps near-optimal alternatives ranked highly, while the margin term orders templates by measured speed with a penalty proportional to the speedup gap. The operator-specific weight $\lambda_o$ controls the strength of the margin term. It is smaller for SpMV, whose template speedups cluster more tightly, and larger for SpMM and SpGEMM.

The additive form decomposes each score into per-feature terms $h_j(x_j)^\mathsf{T}w_{2,o,j,s}$, so the model is interpretable by construction~\cite{agarwal2021nam}. The synthesis stage receives these terms as feature-level explanations for the template ranking. The model returns a ranked set of templates. The ranking therefore guides the known part of the optimization space.
\begin{figure}[t]
  \centering
  \includegraphics[width=\columnwidth]{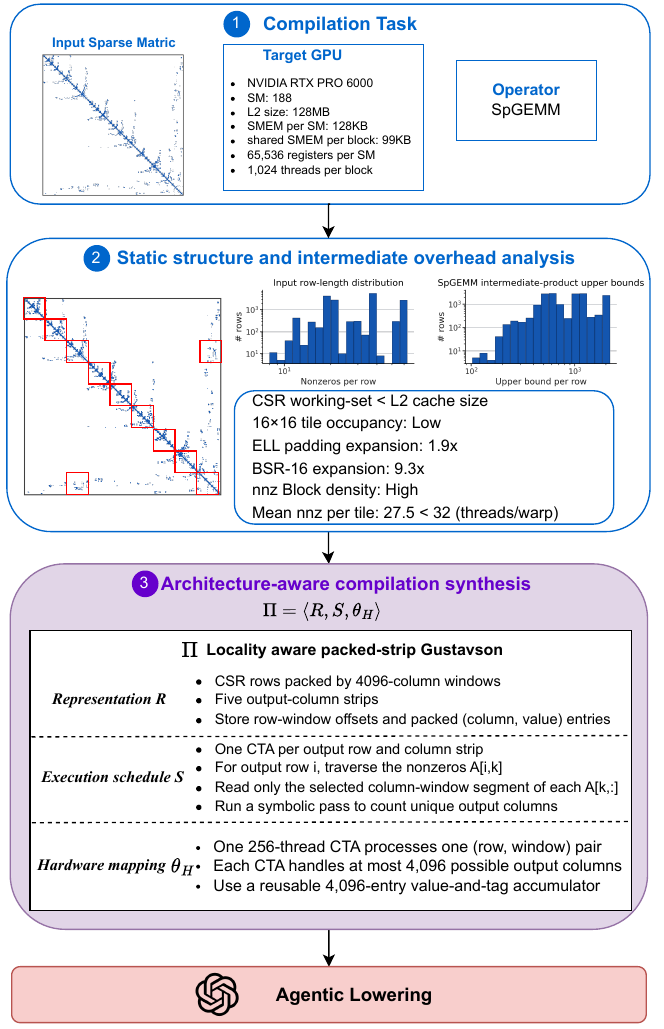}
  \caption{Architecture-aware synthesis of one SpGEMM compilation plan for \texttt{Shen/e40r0100} on an RTX PRO 6000.}
  \label{fig:design-planning}
\end{figure}
\subsection{Architecture-Aware Compilation-Plan Synthesis}
\label{sec:design-planning}

The synthesis stage combines three inputs: the sparse-analysis results, the learned template ranking, and the target-GPU specification, as shown in Figure~\ref{fig:design-overview}. It expands promising templates into complete compilation plans. Each plan records $R$, $S$, and $\theta_H$, together with storage estimates, representation overhead modeling, and on-chip resource bounds.

% Synthesis proceeds in three stages. The first compares workload scale and potential parallelism with physical hardware capacities. The second jointly constructs the algorithm, data representation, and execution schedule. The third binds the hardware mapping to the target GPU.

\noindent\textbf{Workload-scale analysis.}~SparseDitto first estimates the storage and work exposed by a candidate design. Let
\begin{equation}
 M_{\mathrm{rep}}(\Pi)=M_{\mathrm{value}}(L)+M_{\mathrm{meta}}(I)
 \qquad (\mathrm{Bytes})
 \label{eq:representation-footprint}
\end{equation}
be the representation footprint. $M_{\mathrm{value}}$ multiplies the number of stored value slots by the bytes per value. $M_{\mathrm{meta}}$ similarly accounts for the byte width of each metadata field. Let $W(\Pi)$ denote temporary storage introduced by the dataflow, also in bytes. The read-side and total working sets are
\begin{align}
  M_R(\Pi) &= M_{\mathrm{rep}}(\Pi)+W(\Pi)+D_R
    && (\mathrm{Bytes}) \\
  M_T(\Pi) &= M_R(\Pi)+D_S
    && (\mathrm{Bytes})
  \label{eq:candidate-working-set}
\end{align}
where $D_R$ and $D_S$ are the byte footprints of common input and output operands. For SpMM, these terms include dense $B$ and $C$. For SpGEMM, $W(\Pi)$ includes the selected accumulator and intermediate representation.

Let $C_{\mathrm{L2}}$ denote the physical L2 capacity in bytes. The scale ratios are
\begin{equation}
 \begin{aligned}
  q_R(\Pi)&=\frac{M_R(\Pi)}{C_{\mathrm{L2}}},\\
  q_T(\Pi)&=\frac{M_T(\Pi)}{C_{\mathrm{L2}}}.
 \end{aligned}
 \label{eq:l2-scale-ratios}
\end{equation}
SparseDitto also reports the number of independent work units per SM, $N_{\mathrm{work}}(\Pi)/N_{\mathrm{SM}}$. A work unit can be a row, slice, nonzero partition, or tile. These quantities are static scale indicators. They do not predict cache residency or achieved parallelism. Both depend on the lowered code and are determined through target-GPU measurements.

\noindent\textbf{Joint representation-schedule synthesis.}~The second stage uses these estimates and the learned template ranking to construct $R$ and $S$ together. It selects a sparse algorithm and specifies the layout, traversal metadata, parallel decomposition, and dataflow needed to realize it. For example, a SELL plan couples row slicing and slice offsets with a schedule that processes slices in parallel. A SpGEMM hash plan couples row bucketing with the accumulator dataflow. These choices determine both the work assigned to each computing unit and the temporary storage it requires.

%The learned cost model seeds designs from highly ranked templates. The synthesis stage also considers alternatives with different representations or workload decompositions. It updates the storage and work estimates as these decisions are made.

\noindent\textbf{Hardware mapping.}~The third stage binds $R$ and $S$ to target-specific parameters in $\theta_H$, including tile sizes, launch geometry, and on-chip resource allocation. For a hardware mapping $t$, let $B_t$ be its threads per CTA, $r_t$ its registers per thread, and $S_t$ its shared-memory bytes per CTA. Let $T_{\max}$, $R_{\max}$, and $S_{\max}$ denote the corresponding hardware limits. A legal mapping must satisfy
\begin{equation}
 B_t\le T_{\max},\qquad
 r_tB_t\le R_{\max},\qquad
 S_t\le S_{\max}.
 \label{eq:resource-bounds}
\end{equation}

SparseDitto can generate multiple different compilation plan candidates to expand its coverage of the design space. Each completed plan starts an independent search branch. Initial 5 iterations preserve optimization diversity, while later iterations favor the fastest valid plans. Before lowering, the synthesis stage checks the requested thread count and shared-memory allocation. After compilation with \texttt{nvcc}, the reported register count completes the check. 
%Cache and occupancy estimates only guide comparisons among legal plans.

Figure~\ref{fig:design-planning} illustrates the synthesis of one SpGEMM plan for \texttt{Shen/e40r0100} on an RTX PRO 6000. Here, $C=A^2$, $m=n=17{,}281$, and $\operatorname{nnz}(A)=553{,}562$. Scanning the CSR structure gives $\max_i r_i=62$ and $\operatorname{nnzb}_{16}=20{,}130$. Substituting these counts into Equation~\ref{eq:format-expansion} gives $\rho_{\mathrm{ELL}}=17{,}281\times62/553{,}562\approx1.9$ and $\rho_{\mathrm{BSR}}(16)=16^2\times20{,}130/553{,}562\approx9.3$. The same block count gives $553{,}562/20{,}130\approx27.5$ nonzeros per occupied tile, below the 32 threads in a warp; only $20{,}130/\lceil17{,}281/16\rceil^2\approx1.7\%$ of all tiles are occupied. The two histograms show $r_i$ and $u_A(i)=\sum_{k\in A[i,:]}r_k$ from Equation~\ref{eq:bg-spgemm-ub}; the latter ranges from 102 to 2,187 intermediate products per row.

With FP32 values and 32-bit indices, Equation~\ref{eq:representation-footprint} gives the CSR input footprint as $8\times553{,}562+4\times(17{,}281+1)=4{,}497{,}624$ bytes, or 4.29MB. This explains the figure's comparison with the GPU's 128\,MB L2 capacity. The complete plan's working sets and scale ratios additionally include its workspace and other operands through Equations~\ref{eq:candidate-working-set} and~\ref{eq:l2-scale-ratios}.

These estimates guide a packed column-window design that preserves diagonal locality while avoiding dense-block padding. The 4,096-column window and 256-thread CTA are synthesis choices in $\theta_H$. The window width yields 5 output-column strips in $R$, and $S$ assigns one CTA to each row--window pair. 
%A 4,096-entry accumulator with one FP32 value and one 32-bit tag per entry requires $4{,}096\times(4+4)=32{,}768$ bytes (32\,KiB) of shared memory. This allocation is below the 99\,KiB per-CTA limit, and $256\le1{,}024$ satisfies the thread bound in Equation~\ref{eq:resource-bounds}. The resulting plan is passed to agentic lowering.

\subsection{Agentic Lowering and Refinement}
\label{sec:design-codegen}

Each compilation plan is realized as executable CUDA through an agentic lowering stage. The lowering stage receives the selected representation, execution schedule, and hardware mapping from $\Pi$. Its role is to realize these decisions using target-specific CUDA code.

SparseDitto provides conversion tools and metadata builders for CSR, COO, ELL, SELL, and BSR. It also provides adaptive $16{\times}16$ blocks inspired by CB-SpMV~\cite{cong2025cb-spmv}. Packed column windows for SpMM follow Sextans~\cite{sextans}. SpGEMM support includes upper-bound analysis and row binning. Scans and workspace management follow the row-wise structure of MatRaptor~\cite{MatRaptor}. These mechanisms implement $L$ and $I$ while providing building blocks for $P$ and $D$.

An LLM-based coding agent realizes the thread mapping, accumulator design, and $\theta_H$ parameters specified by the plan within a unified harness. Every generated implementation is compiled and numerically validated. SparseDitto then measures latency and profiles individual CUDA kernels with \texttt{NCU} on the target GPU. Profiling feedback is mapped back to the corresponding fields of $\Pi$. Excessive padding or metadata traffic triggers changes to $L$ or $I$. Imbalance and synchronization overhead trigger changes to $P$. Cache behavior and workspace traffic guide changes to $D$. Hardware resource pressure changes $\theta_H$. The synthesis stage updates the compilation plan before the coding agent revises the code. 

SparseDitto returns the fastest numerically valid implementation across all branches. The cached result is identified by its sparsity pattern, operator, and target platform. For SpMM, the key also records $K$. Subsequent invocations with the same key execute the compiled implementation without repeating plan synthesis or lowering.

\section{Evaluation}
\label{sec:eval}

%We evaluate SparseDitto across operators, sparsity patterns, and target GPUs. Profiling and ablations examine the generated implementations and the value of sparse-specific compilation guidance. 
%We also evaluate the learned template ranking, end-to-end application performance, and compilation cost.

% We evaluate whether SparseDitto constructs efficient kernels across
% patterns and operators, whether its selector provides a useful starting
% point, which parts of the workflow contribute to the final result, how
% often the winning kernel leaves the selector vocabulary, and whether
% the kernels hold up inside a real application.

% \subsection{Experimental Setup}
% \label{sec:eval-setup}

\noindent\textbf{Workloads.}~We use a stratified set of $60$ matrices from the SuiteSparse Matrix Collection~\cite{suitesparse}. It covers regular stencils, finite-element matrices, circuit and process-simulation problems, power-law web graphs, road networks, and gene networks, and includes the hardest cases from the evaluations of the specialized systems we compare against. For each matrix, we evaluate SpMV, SpMM with $K\in\{8,32,128,256\}$, and SpGEMM. For SpGEMM, we follow~\cite{wu2025hsmu}, computing $AA$ for square matrices and $AA^\mathsf{T}$ otherwise.

The matrices have 359--16,777,216 rows and 817 to 101 million nonzeros. Their densities range from $1.78\times 10^{-7}$ to 0.110. Mean row length ranges from 1.0 to 1,107.1 nonzeros, and the coefficient of variation of row length ranges from 0 to 12.2.
Appendix~\ref{app:matrices} lists every matrix with these statistics.

\noindent\textbf{Platform and baselines.}~We evaluate on two GPU architectures, NVIDIA RTX PRO 6000 Blackwell with CUDA 13.0 and NVIDIA H200 Hopper with CUDA 12.9. We report the main results on both. The ablation study and the cache and memory-traffic analysis run on the RTX PRO 6000. We use the corresponding cuSPARSE implementation~\cite{cusparse} as the per-task reference. We benchmark all applicable cuSPARSE algorithms under the same numerical contract and use the fastest valid configuration as the baseline. For SpMV, we compare with CB-SpMV~\cite{cong2025cb-spmv} and AlphaSparse~\cite{AlphaSparse}, a auto-tuner that creates machine-designed SpMV formats and kernels. For SpMM, we compare with DTC-SpMM~\cite{dtcspmm} and SparseTIR~\cite{sparsetir}, a sparse compiler for SpMM. For SpGEMM, we compare with HSMU-SpGEMM~\cite{wu2025hsmu}. 

\noindent\textbf{LLM configuration.}~The agents used for plan synthesis, CUDA lowering, and verification call the same model: GPT-5.6-terra~\cite{openai} with reasoning effort \emph{low} and a budget of $8{,}192$ tokens per call. All other parameters are left at the API default. Every task uses the same search budget. Three candidate plans start independent branches, each with up to five lowering-and-evaluation iterations. The two fastest valid branches receive up to five additional iterations for profile-guided refinement.

\noindent\textbf{Measurement protocol.}~We time the generated kernel and baselines with CUDA events. Each implementation has five warmup executions followed by ten timed executions, and we report their arithmetic mean, excluding the one-time preprocessing. Aggregate speedups are geometric means over tasks, reported with a $95\%$ confidence interval obtained by bootstrapping the tasks ($20{,}000$ resamples of the log-speedups).
% . One-time preprocessing is excluded from the reported kernel time. 
%Unsupported $K$ for SparseTIR and unsupported matrices for cuSPARSE on SpGEMM are excluded.

\noindent\textbf{Correctness validation.}~We use the corresponding FP32 cuSPARSE implementation as per-task correctness reference. SparseDitto and cuSPARSE receive identical input values. For SpMV and SpMM, a candidate is considered correct only if every output entry satisfies
\begin{equation}
\left|\widehat{y}_i - y_i^{\mathrm{ref}}\right|
\leq
\epsilon_{\mathrm{abs}}
+
\epsilon_{\mathrm{rel}}
\left|y_i^{\mathrm{ref}}\right|,
\end{equation}
where $\epsilon_{\mathrm{abs}} = 10^{-5}$ and
$\epsilon_{\mathrm{rel}} = 10^{-3}$. Any runtime error, unexpected
\texttt{NaN} or infinity, or violation of this bound causes the candidate to fail validation.

For SpGEMM, correctness includes both the sparse output structure and the associated values. We canonicalize the SparseDitto and cuSPARSE outputs by sorting column indices within each row and coalescing duplicate entries. The output dimensions, number of nonzeros, row pointers, and column indices must match exactly. Every corresponding nonzero value must additionally satisfy the same mixed absolute and relative error bound above. We apply this single predeclared validation rule uniformly to all matrices, generated candidates, and comparison systems.

\begin{figure}[t]
  \centering
  \includegraphics[width=0.92\columnwidth,trim=0 7bp 0 8bp,clip]{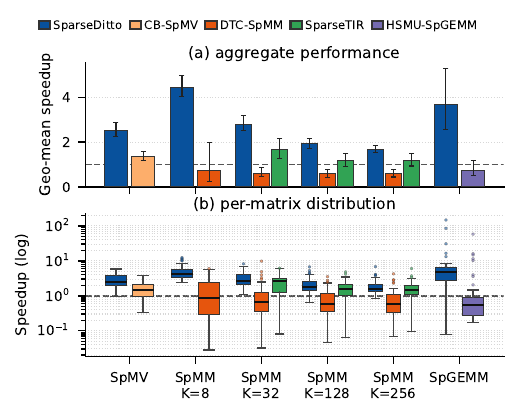}
  \caption{Generated kernel performance for SpMV, SpMM, and SpGEMM. (a)
  geometric-mean speedup over cuSPARSE; whiskers are $95\%$ bootstrap
  confidence intervals over tasks. (b) the per-matrix distribution
  on a logarithmic scale. The dashed line is cuSPARSE.}
  \label{fig:eval-operators}
\end{figure}
\begin{figure}[t]
  \centering
  \includegraphics[width=0.92\columnwidth,trim=0 7bp 0 8bp,clip]{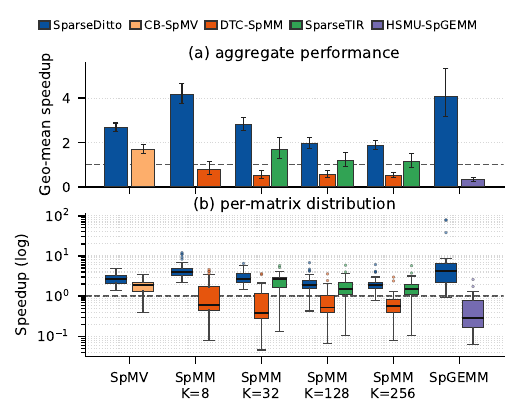}
  \caption{The same measurement on the H200. Both panels follow Figure~\ref{fig:eval-operators}.}
  \label{fig:eval-operators-h200}
\end{figure}

\subsection{Generated Kernel Performance}
\label{sec:eval-performance}

We organize the results by operator and report each one on both GPUs. On seven of the largest matrices, the cuSPARSE SpGEMM implementation triggers an out-of-resources error, so those tasks have no reference and are excluded. Each additional system is evaluated only for the operator and configurations supported by its released artifact. Aggregate speedups are geometric means over the validated tasks.

On the RTX PRO 6000, SparseDitto reaches $2.68\times$ over cuSPARSE ($95\%$ CI $[2.48,2.89]$). On the H200, the same compilation framework reads the new target's device properties and regenerates every implementation from scratch. It reaches $2.79\times$ ($95\%$ CI $[2.61,2.99]$). Figures~\ref{fig:eval-operators} and~\ref{fig:eval-operators-h200} break down the results by operator and dense width.

\noindent\textbf{SpMV.}~Against cuSPARSE, SparseDitto achieves $2.53\times$ on the RTX PRO 6000 and $2.68\times$ on the H200. CB-SpMV achieves $1.35\times$ and $1.70\times$ on the two GPUs. On the same inputs, SparseDitto is $1.91\times$ and $1.57\times$ faster than CB-SpMV.

AlphaSparse generates a customized SpMV implementation for each matrix by searching a design space defined by static rules. With a nine-hour search budget per matrix, it produces a kernel that passes our validation rule on 13 matrices. The remaining runs crash or return values outside the error bound. On those 13 matrices, SparseDitto is $5.34\times$ faster than AlphaSparse.

\noindent\textbf{SpMM.}~Against cuSPARSE, SparseDitto achieves $4.46\times$ at $K=8$, $2.81\times$ at $K=32$, $1.92\times$ at $K=128$, and $1.68\times$ at $K=256$. The gain is largest at the small dense widths and narrows as $K$ grows. The H200 follows the same trend, at $4.18\times$, $2.81\times$, $1.96\times$, and $1.87\times$.

For SpMM, we also compare with SparseTIR and DTC-SpMM. SparseTIR reaches $1.68\times$ at $K=32$. Its speedup is $1.18\times$ at both $K=128$ and $K=256$. DTC-SpMM achieves $0.73\times$ at $K=8$ and $0.62\times$, $0.58\times$, and $0.58\times$ at the three larger widths, so it stays below cuSPARSE at every width. Across the evaluated tasks, SparseDitto is $1.57\times$ faster than SparseTIR and $3.99\times$ faster than DTC-SpMM. On the H200, the corresponding speedups are $1.63\times$ and $4.32\times$.

\noindent\textbf{SpGEMM.}~SparseDitto achieves $3.70\times$ over cuSPARSE on the RTX PRO 6000 and $4.09\times$ on the H200 (up to $146.61\times$). HSMU-SpGEMM achieves $0.74\times$ on the square matrices, and $0.32\times$ on the H200. On these matched inputs, SparseDitto is $5.37\times$ faster than HSMU-SpGEMM on the RTX PRO 6000 and $13.60\times$ faster on the H200. SparseDitto also covers the rectangular inputs that HSMU-SpGEMM does not support.

Panel (b) of both figures reports the per-matrix distribution. At $K=8$ and $K=32$, the whole SpMM distribution sits above cuSPARSE on both GPUs. As $K$ grows, the box contracts toward the baseline. On the RTX PRO 6000, the interquartile range narrows from $3.28$--$5.57\times$ at $K=8$ to $1.27$--$2.22\times$ at $K=256$. SpGEMM has the widest spread on both platforms, with maxima of $146.61\times$ and $78.5\times$. This variation is consistent with its sensitivity to intermediate-product distributions and accumulator design, beyond the input nonzero count alone.
% The lower tails locate the losses, and they are shorter on the H200: SpMV loses on three matrices there against none, and SpGEMM on eight against four.

\subsection{Cache and Memory-Traffic Analysis}
\label{sec:eval-cache}
We use NVIDIA Nsight Compute to examine how the generated implementations use the memory hierarchy and GPU resources. We profile the timed kernels of each winning implementation, its cuSPARSE reference, and the specialized systems on the same inputs. The profiler replays each launch with flushed caches and locked base clocks. When a compute phase launches several kernels, we pool the sector counters over all launches. DRAM traffic is measured in bytes read and written at device memory. Table~\ref{tab:eval-cache} summarizes cache hit rates and request counts, together with achieved occupancy and DRAM bandwidth.
\begin{table}[h]
  \caption{SparseDitto against each system on evaluated tasks. $\Delta$L1 and $\Delta$L2 are our hit rate minus theirs, in percentage points. The remaining columns are ours divided by theirs: L1 and L2 sector requests per operator invocation, achieved occupancy, and achieved DRAM bandwidth.}
  \label{tab:eval-cache}
  \centering \footnotesize \setlength{\tabcolsep}{4pt}
  \begin{tabular}{@{}l r r r r r r@{}}
    \toprule
    \textbf{Comparison}
    & \textbf{$\Delta$L1} & \textbf{L1 req.}
    & \textbf{$\Delta$L2} & \textbf{L2 req.}
    & \textbf{Occ.} & \textbf{BW} \\
    \midrule
    vs.\ cuSPARSE    & $+4.7$ & $1.09\times$ & $+3.6$ & $0.77\times$ & $1.96\times$ & $1.98\times$ \\
    vs.\ CB-SpMV     & $+1.6$ & $0.63\times$ & $-7.3$ & $0.57\times$ & $0.93\times$ & $1.03\times$ \\
    vs.\ SparseTIR   & $-18.6$ & $0.85\times$ & $-2.2$ & $1.03\times$ & $1.24\times$ & $1.80\times$ \\
    vs.\ DTC-SpMM    & $-5.1$ & $1.44\times$ & $-1.3$ & $1.13\times$ & $3.21\times$ & $5.03\times$ \\
    vs.\ HSMU-SpGEMM & $-16.3$ & $0.83\times$ & $+1.5$ & $0.97\times$ & $0.88\times$ & $4.25\times$ \\
    \bottomrule
  \end{tabular}
\end{table}

\noindent\textbf{Versus cuSPARSE.}~SparseDitto issues slightly more L1 requests but fewer L2 requests, with higher hit rates at both levels. Its achieved occupancy and DRAM bandwidth are both about twice those of cuSPARSE. The gains thus accompany improvements in both memory access behavior and hardware utilization. The cache effects differ by operator. For SpMM, the median L1 hit-rate increase ranges from $8$ to $17$ percentage points across dense widths. For SpGEMM, the improvement appears primarily in L2. 
%These operator-dependent cache effects support adapting the dataflow to each workload, alongside the hardware mapping.
%SpMV's hit rates remain within a few points of cuSPARSE, so higher hit rates alone do not explain its speedup.

\noindent\textbf{Versus specialized systems.}~SparseTIR and HSMU-SpGEMM have higher L1 hit rates than SparseDitto, yet SparseDitto is $1.57\times$ and $5.37\times$ faster on the same inputs. Against SparseTIR, SparseDitto issues $0.85\times$ as many L1 requests and $1.03\times$ as many L2 requests, while sustaining $1.80\times$ the DRAM bandwidth. Against CB-SpMV, it reduces L1 and L2 requests to $0.63\times$ and $0.57\times$, despite a lower L2 hit rate.
%Higher hit rates need not imply less work or better hardware utilization.

Against HSMU-SpGEMM, SparseDitto achieves $0.88\times$ the occupancy but $4.25\times$ the DRAM bandwidth. The profiled execution uses $5$ kernels per SpGEMM, compared with $24$ for HSMU-SpGEMM. Fewer stages reduce launch overhead and the number of inter-kernel dependencies. This comparison highlights the importance of the accumulation dataflow and kernel decomposition, beyond cache hit rates or occupancy.

DTC-SpMM illustrates the interaction between representation overhead and hardware mapping. It moves slightly less data than SparseDitto, yet SparseDitto is $3.99\times$ faster. SparseDitto achieves $3.21\times$ its occupancy and $5.03\times$ its DRAM bandwidth. DTC-SpMM's Tensor-Core pipeline reaches only $1.5\%$ of peak, and its $16\times 8$ tiles have a median nonzero occupancy of $10.1\%$ on these matrices. This low tile occupancy leaves much of the blocked computation operating on padding. Low achieved warp occupancy further limits latency hiding. SparseDitto's advantage decreases from $6.40\times$ at $K=8$ to $2.89\times$ at $K=256$ as the dense width grows.

Together, these comparions reveal that higher cache hit rates or occupancy alone do not gurantee better performance. SparseDitto outperms baselines by covering broader optimization space insead of focusing on any single metrics. 

%These comparisons show why a compilation plan should coordinate representation, execution schedule, and hardware mapping. Target-GPU profiling reveals how these choices affect request volume and useful computation in the lowered code, providing feedback for plan refinement.

\subsection{Ablation Study and Comparison with General Agentic CUDA Coding Framework}
\label{sec:eval-ablation}

We firstly compare with one SOTA general agentic CUDA coding framework StitchCUDA~\cite{stitchcuda}, whose workflow and prompts design are tailored for general CUDA tasks. It achieves $1.53\times$, $1.76\times$, and $0.86\times$ over cuSPARSE on three operators seperately. Then we evaluate three configurations with progressively richer sparse-specific guidance. \emph{Learned Model + Agentic Lowering} generates implementations with guide from our learned cost model. The third augments the task with static pattern analysis as input to agentic lowering stage. The full configuration adds architecture-aware synthesis of compilation plans that coordinate representation, schedule, and hardware mapping.
% We evaluate three configurations under the same generation budget.
% \emph{Task only} receives the task description and cuSPARSE reference.
% It follows one optimization trajectory without structural analysis.
% \emph{Pattern + search} adds sparsity-pattern features and independent
% search branches. It omits the learned selector and the static workload
% bounds. The full configuration adds strategy selection and
% architecture-aware planning.

Table~\ref{tab:eval-ablation} reports geometric-mean speedups for each configuration. \emph{Learned Model + Agentic Lowering} already outperforms cuSPARSE on average. \emph{Pattern Analysis} raises SpMV speedup from $1.93\times$ to $2.27\times$ and SpMM speedup from $2.02\times$ to $2.28\times$. The larger gain is on SpGEMM, from $1.22\times$ to $2.24\times$. The full configuration further improves all three operators, reaching $2.53\times$, $2.52\times$, and $3.70\times$, respectively. This progression supports sparse-specific guidance and architecture-aware plan synthesis significantly contribute to generated kernels' performance.
\begin{table}[h!t]
  \caption{Ablation results by operator. Each cell reports the
  geometric-mean speedup over cuSPARSE.}
  \label{tab:eval-ablation}
  \centering
  \footnotesize
  \setlength{\tabcolsep}{5pt}
  \begin{tabular}{@{}l r r r@{}}
    \toprule
    \textbf{Configuration}
    & \textbf{SpMV}
    & \textbf{SpMM}
    & \textbf{SpGEMM} \\
    \midrule
    StitchCUDA~\cite{stitchcuda} & $1.53\times$ &$1.76\times$ &$0.86\times$ \\
    Learned Model + Agentic Lowering
      & $1.93\times$ & $2.02\times$ & $1.22\times$ \\
    * + Pattern Analysis
      & $2.27\times$ & $2.28\times$ & $2.24\times$ \\
    * + Architecture-aware Synthesis (full)
      & $\mathbf{2.53\times}$
      & $\mathbf{2.52\times}$
      & $\mathbf{3.70\times}$ \\
    \bottomrule
  \end{tabular}
\end{table}
%Pattern + Task raises SpMV from $2.48\times$ to $2.64\times$ and SpMM from $2.17\times$ to $2.46\times$. Its larger effect on SpGEMM raises the speedup from $1.25\times$ to $2.29\times$. The full system further improves the three operators to $2.95\times$, $2.60\times$, and $3.78\times$. This consistent progression shows that each stage improves the generated kernels under the same generation budget.

%The larger improvement on SpGEMM is consistent with its additional dependence on intermediate-product distributions, output construction, and accumulator organization. These results support the effectiveness of the integrated sparse compilation pipeline beyond task-conditioned and feature-augmented generation. 

\subsection{Learned Cost Model}
\label{sec:eval-selector}

We next evaluate the learned cost model's ability to rank optimization templates and the structural features that influence its ranking. This ranking supplies a prior for compilation-plan synthesis; the final implementation is selected through target-GPU measurements.

\noindent\textbf{Training cost.}~The additive energy model has $15.9$K parameters. Training on the full corpus of $400$ matrices for $600$ epochs takes about two seconds on a single RTX PRO 6000 or H200 GPU. At inference, ranking the templates takes well under a microsecond, a negligible cost relative to plan synthesis and agentic CUDA lowering.

\noindent\textbf{Ranking quality.}~Our corpus contains $400$ SuiteSparse matrices with measured template runtimes. It does not overlap with the evaluation set in Section~\ref{sec:eval-performance}. It also outperformes multilayer perceptron or gradient-boosted trees~\cite{friedman2001greedy}, while retaining an additive score decomposition. Full protocols, metric definitions, and per-operator results are provided in Appendix~\ref{app:selector}.
%Compared with a multilayer perceptron trained with the same listwise objective, the energy model improves ranking quality on SpMM and SpGEMM and achieves comparable results on SpMV. It also remains competitive with gradient-boosted trees~\cite{friedman2001greedy},  

\begin{figure}[t]
  \centering
  \includegraphics[width=\columnwidth,trim=0 10bp 0 7bp,clip]{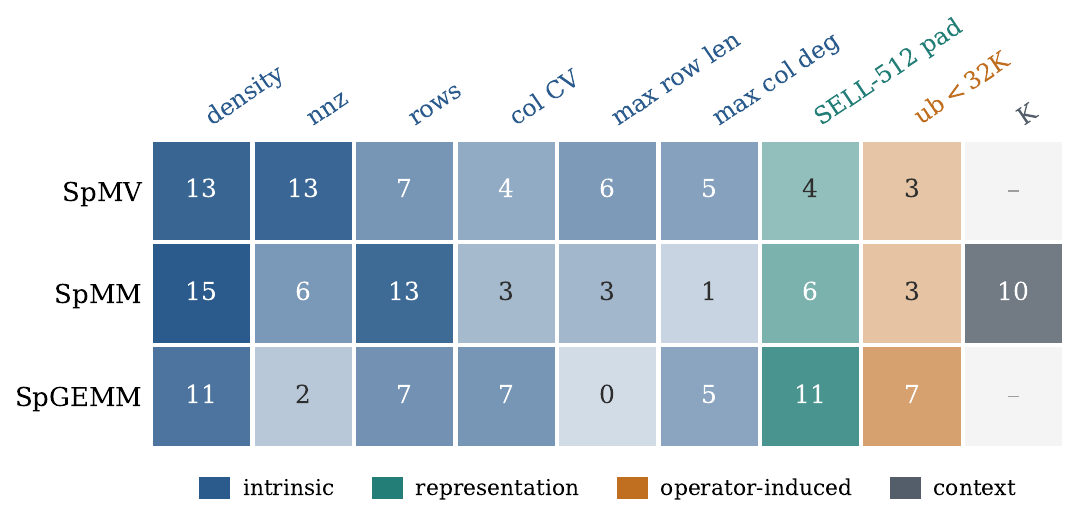}
  \caption{Structural feature importance per operator. Each cell gives a
  feature's share of permutation importance in percent; colors group the
  feature families, and a dash marks a feature the operator does not use.}
  \label{fig:feat-importance}
\end{figure}

\noindent\textbf{Which features matter.}~Figure~\ref{fig:feat-importance} reports permutation importance, which measures the model's sensitivity to perturbing each feature across inputs. Matrix density, nonzero count, and row count are influential across all three operators. Operator-specific features also matter: the dense width $K$ for SpMM and intermediate-product upper bounds for SpGEMM. These results support using both intrinsic pattern features and operator-dependent work estimates to rank templates. For an individual input, the additive score decomposition provides feature-level guidance to synthesis. Appendix~\ref{app:importance} reports the complementary model-native importance measure.

\subsection{End-to-End Application Evaluation}
\label{sec:eval-gnn}

We evaluate whether SparseDitto's specialized implementations improve application performance in a two-layer full-batch GCN on Reddit~\cite{reddit}, following DTC-SpMM's protocol~\cite{dtcspmm}. We use an unweighted symmetric adjacency and sweep $K\in\{8,32,128,256\}$. For each graph and dense width, SparseDitto generates the aggregation kernel once. The kernel can be reused across epochs and training runs on the target GPU. Changes to node features and model parameters do not require regenerating it. Each result averages $100$ epochs after $10$ warmup epochs. Each epoch invokes four sparse aggregations: two in the forward pass and two in the backward pass. Symmetry also lets the backward pass reuse the corresponding forward kernel. Only the sparse aggregation implementation is replaced; the rest of the training loop is unchanged.

\begin{figure}[t]
  \centering
  \includegraphics[width=0.93\columnwidth,trim=0 7bp 0 8bp,clip]{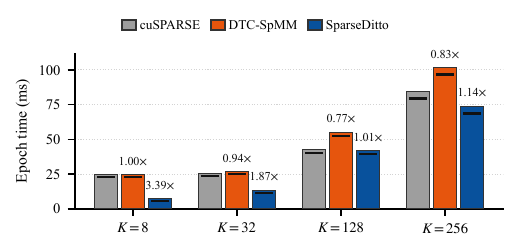}
  \caption{Full-batch GCN epoch time on Reddit across hidden widths.
  Horizontal marks inside bars denote SpMM time. Labels above DTC-SpMM
  and SparseDitto report speedup over cuSPARSE.}
  \Description{A grouped bar chart compares epoch time for cuSPARSE,
  DTC-SpMM, and SparseDitto at hidden widths 8, 32, 128, and 256.
  SparseDitto has the lowest epoch time at every width.}
  \label{fig:eval-gnn}
\end{figure}

Figure~\ref{fig:eval-gnn} reports the result. At $K=8$, SparseDitto reduces the average epoch time from $24.6$ to $7.3$ ms, yielding a $3.39\times$ end-to-end speedup over cuSPARSE. At $K=32$, it reduces the epoch from $25.2$ to $13.5$ ms and achieves $1.87\times$. The advantage narrows as the dense dimension grows, but SparseDitto remains faster than cuSPARSE. It reaches $1.01\times$ at $K=128$ and $1.14\times$ at $K=256$. DTC-SpMM matches cuSPARSE only at $K=8$. Its speedups fall to $0.94\times$, $0.77\times$, and $0.83\times$ at the other widths. SparseDitto is therefore $1.31\times$--$3.38\times$ faster than DTC-SpMM across the sweep.

% The breakdown inside each bar explains how kernel performance carries to the application. SpMM accounts for $77\%$--$95\%$ of epoch time. At $K=8$, SparseDitto accelerates the aggregation part from $22.9$ to $5.6$ ms, a $4.09\times$ kernel-level improvement. The fixed GCN work outside SpMM reduces the end-to-end gain to $3.39\times$. The same trend holds at the other widths: the SpMM speedups are $2.02\times$, $1.01\times$, and $1.16\times$, closely tracking the corresponding epoch speedups. After subtracting SpMM time, the remaining epoch cost is nearly identical across the three systems at each $K$. It ranges from $1.7$ to $5.2$ ms. This isolates the measured improvement to the generated aggregation kernels rather than unrelated training work.

% We also verify correctness at the application level. In a lockstep Reddit run from identical initialization, the loss trajectory using SparseDitto stays within $6\times10^{-6}$ relative deviation of cuSPARSE over twelve epochs. This difference is consistent with FP32 reduction-order effects. The generated kernels move from the benchmark harness into the training loop without modification. Their one-time preprocessing cost remains below one millisecond and is excluded from the timed epochs.

\subsection{Compilation and Lowering Cost}
\label{sec:eval-search}

We report the one-time cost of constructing a specialized implementation, including plan synthesis, lowering, and target-GPU evaluation. SparseDitto takes about $30$ minutes per matrix on average under the generation budget described above. LLM inference dominates this compilation time. Each task runs at most a few dozen compile-and-measure cycles, with each cycle taking seconds to tens of seconds depending on matrix size. The API cost for the main evaluation is about \$20 per matrix. These construction costs are separate from the execution times reported in Section~\ref{sec:eval-performance}.

AlphaSparse~\cite{AlphaSparse} similarly treats specialized kernel generation as an offline construction task. We allow its SpMV search a nine-hour budget per matrix. On \texttt{Gset/G31}, a graph matrix with $99\%$ sparsity, it compiles and runs $730$ candidate kernels within that budget. Its best kernel reaches $1.35\times$ over cuSPARSE. SparseDitto reaches $4.46\times$ on the same matrix in about $30$ minutes.

%\subsection{Compilation Framework Reusability}
%SparseDitto retains the compiled implementation for reuse with the similar sparsity pattern, operator, and target platform. Subsequent invocations reuse this result without repeating plan synthesis or lowering. We also develop high-performance parrallel algorithm for matrix pattern anaylsis on GPU, making the on-the-fly pattern analysis only takes very minor time, e.g., 10 - 30 ms, depending on matrix size. The extra overhead incurred by traditional sparse compilers, e.g., sparseTIR, can be up to hundreds of ms. The detail of this evaluation  is in supplementary material.

\section{Conclusion}
\label{sec:conclusion}

The performance of sparse matrix operators on GPU depends jointly on the data representation, the execution schedule, and the hardware mapping, which need to all match the sparsity pattern. SparseDitto addresses this challenge with a unified compilation framework, covering SpMV, SpMM, and SpGEMM. Within this framework, an interpretable energy model ranks established strategy templates from structural features, an architecture-aware synthesis stage explores the design space under target-GPU constraints, and LLM agents automatically lowering this plan into CUDA code. SparseDitto outperforms cuSPARSE in evaluated tasks, achieving geometric-mean speedups of 2.68$\times$ on an RTX PRO 6000 and 2.79$\times$ on an H200. It runs 1.57$\times$ to 5.37$\times$ faster than four state-of-the-art specialized systems and accelerates full-batch GCN training by up to 3.39$\times$.

\clearpage
\bibliographystyle{ACM-Reference-Format}
\bibliography{sample,drsparse}

@inproceedings{cong2025cb-spmv,
  title={CB-SpMV: A Data Aggregating and Balance Algorithm for for Cache-Friendly Block-Based SpMV on GPUs},
  author={Cong, Xing and Sun, FuKai and Chen, YiFan and Xie, Chenhao and Liu, Yi and Qian, Depei},
  booktitle={Proceedings of the 39th ACM International Conference on Supercomputing},
  pages={149--160},
  year={2025}
}

@inproceedings{wu2025hsmu,
  title={HSMU-SpGEMM: Achieving High Shared Memory Utilization for Parallel Sparse General Matrix-Matrix Multiplication on Modern GPUs},
  author={Wu, Min and Luo, Huizhang and Li, Fenfang and Zhang, Yiran and Tang, Zhuo and Li, Kenli and Zhang, Jeff and Liu, Chubo},
  booktitle={2025 IEEE International Symposium on High Performance Computer Architecture (HPCA)},
  pages={1452--1466},
  year={2025},
  organization={IEEE}
}

@inproceedings{TC-GNN,
  title={TC-GNN: Bridging Sparse GNN Computation and Dense Tensor Cores on GPUs},
  author={Yuke Wang and Boyuan Feng and Zheng Wang and Guyue Huang and Yufei Ding},
  booktitle={USENIX Annual Technical Conference (ATC)},
  year={2023}
}

@article{qimeng,
      title={QiMeng-Kernel: Macro-Thinking Micro-Coding Paradigm for LLM-Based High-Performance GPU Kernel Generation},
  author={Zhu, Xinguo and Peng, Shaohui and Guo, Jiaming and Chen, Yunji and Guo, Qi and Wen, Yuanbo and Qin, Hang and Chen, Ruizhi and Zhou, Qirui and Gao, Ke and others},
  journal={arXiv preprint arXiv:2511.20100},
  year={2025}
}

@misc{kernelbench,
      title={KernelBench: Can LLMs Write Efficient GPU Kernels?}, 
      author={Anne Ouyang and Simon Guo and Simran Arora and Alex L. Zhang and William Hu and Christopher Ré and Azalia Mirhoseini},
      year={2025},
      eprint={2502.10517},
      archivePrefix={arXiv},
      primaryClass={cs.LG},
      url={https://arxiv.org/abs/2502.10517}, 
}

@inproceedings{stitchcuda,
  title     = {StitchCUDA: An Automated Multi-Agents End-to-End GPU Programming Framework with Rubric-based Agentic Reinforcement Learning},
  author    = {Li, Shiyang and Zhang, Zijian and Chen, Winson and Luo, Yuebo and Hong, Mingyi and Ding, Caiwen},
  booktitle = {Proceedings of the 43rd International Conference on Machine Learning},
  year      = {2026},
  series = {ICML'26}
}

@inproceedings{TileSpGEMM,
author = {Niu, Yuyao and Lu, Zhengyang and Ji, Haonan and Song, Shuhui and Jin, Zhou and Liu, Weifeng},
title = {TileSpGEMM: a tiled algorithm for parallel sparse general matrix-matrix multiplication on GPUs},
year = {2022},
isbn = {9781450392044},
publisher = {Association for Computing Machinery},
address = {New York, NY, USA},
url = {https://doi.org/10.1145/3503221.3508431},
doi = {10.1145/3503221.3508431},
booktitle = {Proceedings of the 27th ACM SIGPLAN Symposium on Principles and Practice of Parallel Programming},
pages = {90–106},
numpages = {17},
location = {Seoul, Republic of Korea},
series = {PPoPP '22}
}

@inproceedings{AlphaSparse,
author = {Du, Zhen and Li, Jiajia and Wang, Yinshan and Li, Xueqi and Tan, Guangming and Sun, Ninghui},
title = {AlphaSparse: generating high performance SpMV codes directly from sparse matrices},
year = {2022},
isbn = {9784665454445},
publisher = {IEEE Press},
booktitle = {Proceedings of the International Conference on High Performance Computing, Networking, Storage and Analysis},
articleno = {66},
numpages = {15},
location = {Dallas, Texas},
series = {SC '22}
}

@article{suitesparse,
author = {Davis, Timothy A. and Hu, Yifan},
title = {The University of Florida Sparse Matrix Collection},
journal = {ACM Transactions on Mathematical Software},
volume = {38},
number = {1},
year = {2011},
pages = {1:1--1:25},
publisher = {Association for Computing Machinery}
}

@misc{cusparse,
author = {{NVIDIA Corporation}},
title = {{cuSPARSE} Library Documentation},
year = {2026},
url = {https://docs.nvidia.com/cuda/cusparse/}
}

@article{sellcsigma,
author = {Kreutzer, Moritz and Hager, Georg and Wellein, Gerhard and Fehske, Holger and Bishop, Alan R.},
title = {A Unified Sparse Matrix Data Format for Efficient General Sparse Matrix-Vector Multiplication on
Modern Processors with Wide SIMD Units},
year = {2014},
issue_date = {2014},
publisher = {Society for Industrial and Applied Mathematics},
address = {USA},
volume = {36},
number = {5},
issn = {1064-8275},
url = {https://doi.org/10.1137/130930352},
doi = {10.1137/130930352},
month = jan,
pages = {C401–C423},
numpages = {23}
}

@inproceedings{mergespmv,
  author={Merrill, Duane and Garland, Michael},
  booktitle={SC '16: Proceedings of the International Conference for High Performance Computing, Networking, Storage and Analysis}, 
  title={Merge-Based Parallel Sparse Matrix-Vector Multiplication}, 
  year={2016},
  volume={},
  number={},
  pages={678-689},
  doi={10.1109/SC.2016.57}
}

@inproceedings{dtcspmm,
author = {Fan, Ruibo and Wang, Wei and Chu, Xiaowen},
title = {DTC-SpMM: Bridging the Gap in Accelerating General Sparse Matrix Multiplication with Tensor Cores},
year = {2024},
isbn = {9798400703867},
publisher = {Association for Computing Machinery},
address = {New York, NY, USA},
url = {https://doi.org/10.1145/3620666.3651378},
doi = {10.1145/3620666.3651378},
booktitle = {Proceedings of the 29th ACM International Conference on Architectural Support for Programming Languages and Operating Systems, Volume 3},
pages = {253–267},
numpages = {15},
location = {La Jolla, CA, USA},
series = {ASPLOS '24}
}

@inproceedings{nsparse,
  author={Nagasaka, Yusuke and Nukada, Akira and Matsuoka, Satoshi},
  booktitle={2017 46th International Conference on Parallel Processing (ICPP)}, 
  title={High-Performance and Memory-Saving Sparse General Matrix-Matrix Multiplication for NVIDIA Pascal GPU}, 
  year={2017},
  volume={},
  number={},
  pages={101-110},
  doi={10.1109/ICPP.2017.19}
}

@inproceedings{speck,
author = {Parger, Mathias and Winter, Martin and Mlakar, Daniel and Steinberger, Markus},
title = {spECK: accelerating GPU sparse matrix-matrix multiplication through lightweight analysis},
year = {2020},
isbn = {9781450368186},
publisher = {Association for Computing Machinery},
address = {New York, NY, USA},
url = {https://doi.org/10.1145/3332466.3374521},
doi = {10.1145/3332466.3374521},
pages = {362–375},
numpages = {14},
location = {San Diego, California},
series = {PPoPP '20}
}

@article{taco,
author = {Kjolstad, Fredrik and Kamil, Shoaib and Chou, Stephen and Lugato, David and Amarasinghe, Saman},
title = {The tensor algebra compiler},
year = {2017},
issue_date = {October 2017},
publisher = {Association for Computing Machinery},
address = {New York, NY, USA},
volume = {1},
number = {OOPSLA},
url = {https://doi.org/10.1145/3133901},
doi = {10.1145/3133901},
journal = {Proc. ACM Program. Lang.},
month = oct,
articleno = {77},
numpages = {29}
}

@inproceedings{sparsetir,
author = {Ye, Zihao and Lai, Ruihang and Shao, Junru and Chen, Tianqi and Ceze, Luis},
title = {SparseTIR: Composable Abstractions for Sparse Compilation in Deep Learning},
year = {2023},
isbn = {9781450399180},
publisher = {Association for Computing Machinery},
address = {New York, NY, USA},
url = {https://doi.org/10.1145/3582016.3582047},
doi = {10.1145/3582016.3582047},
booktitle = {Proceedings of the 28th ACM International Conference on Architectural Support for Programming Languages and Operating Systems, Volume 3},
pages = {660–678},
numpages = {19},
location = {Vancouver, BC, Canada},
series = {ASPLOS 2023}
}

@article{coo,
author = {Filippone, Salvatore and Cardellini, Valeria and Barbieri, Davide and Fanfarillo, Alessandro},
title = {Sparse Matrix-Vector Multiplication on GPGPUs},
year = {2017},
issue_date = {December 2017},
publisher = {Association for Computing Machinery},
address = {New York, NY, USA},
volume = {43},
number = {4},
issn = {0098-3500},
url = {https://doi.org/10.1145/3017994},
doi = {10.1145/3017994},
journal = {ACM Trans. Math. Softw.},
month = jan,
articleno = {30},
numpages = {49}
}

@inproceedings{csr5,
author = {Liu, Weifeng and Vinter, Brian},
title = {CSR5: An Efficient Storage Format for Cross-Platform Sparse Matrix-Vector Multiplication},
year = {2015},
isbn = {9781450335591},
publisher = {Association for Computing Machinery},
address = {New York, NY, USA},
url = {https://doi.org/10.1145/2751205.2751209},
doi = {10.1145/2751205.2751209},
booktitle = {Proceedings of the 29th ACM on International Conference on Supercomputing},
pages = {339–350},
numpages = {12},
location = {Newport Beach, California, USA},
series = {ICS '15}
}

@inproceedings{MatRaptor,
  author={Srivastava, Nitish and Jin, Hanchen and Liu, Jie and Albonesi, David and Zhang, Zhiru},
  booktitle={2020 53rd Annual IEEE/ACM International Symposium on Microarchitecture (MICRO)},
  title={MatRaptor: A Sparse-Sparse Matrix Multiplication Accelerator Based on Row-Wise Product},
  year={2020},
  volume={},
  number={},
  pages={766-780},
  doi={10.1109/MICRO50266.2020.00068}
}

@inproceedings{sextans,
author = {Song, Linghao and Chi, Yuze and Sohrabizadeh, Atefeh and Choi, Young-kyu and Lau, Jason and Cong, Jason},
title = {Sextans: A Streaming Accelerator for General-Purpose Sparse-Matrix Dense-Matrix Multiplication},
year = {2022},
isbn = {9781450391498},
publisher = {Association for Computing Machinery},
address = {New York, NY, USA},
url = {https://doi.org/10.1145/3490422.3502357},
doi = {10.1145/3490422.3502357},
booktitle = {Proceedings of the 2022 ACM/SIGDA International Symposium on Field-Programmable Gate Arrays},
pages = {65–77},
numpages = {13},
location = {Virtual Event, USA},
series = {FPGA '22}
}

@incollection{lecun2006tutorial,
  title={A tutorial on energy-based learning},
  author={LeCun, Yann and Chopra, Sumit and Hadsell, Raia and Ranzato, Marc'Aurelio and Huang, Fu Jie},
  booktitle={Predicting Structured Data}, publisher={MIT Press}, year={2006}}

@book{hastie1990gam,
  title={Generalized Additive Models},
  author={Hastie, Trevor J. and Tibshirani, Robert J.},
  publisher={Chapman \& Hall/CRC}, year={1990}}

@inproceedings{agarwal2021nam,
  title={Neural additive models: Interpretable machine learning with neural nets},
  author={Agarwal, Rishabh and Melnick, Levi and Frosst, Nicholas and Zhang, Xuezhou and Lengerich, Ben and Caruana, Rich and Hinton, Geoffrey E.},
  booktitle={Advances in Neural Information Processing Systems (NeurIPS)}, volume={34}, pages={4699--4711}, year={2021}}

@inproceedings{burges2005ranknet,
  title={Learning to rank using gradient descent},
  author={Burges, Chris and Shaked, Tal and Renshaw, Erin and Lazier, Ari and Deeds, Matt and Hamilton, Nicole and Hullender, Greg},
  booktitle={Proc. 22nd Int. Conf. on Machine Learning (ICML)}, pages={89--96}, year={2005}}

@inproceedings{cao2007listnet,
  title={Learning to rank: from pairwise approach to listwise approach},
  author={Cao, Zhe and Qin, Tao and Liu, Tie-Yan and Tsai, Ming-Feng and Li, Hang},
  booktitle={Proc. 24th Int. Conf. on Machine Learning (ICML)}, pages={129--136}, year={2007}}

@inproceedings{sedaghati2015automatic,
  title={Automatic selection of sparse matrix representation on GPUs},
  author={Sedaghati, Naser and Mu, Te and Pouchet, Louis-No{\"e}l and Parthasarathy, Srinivasan and Sadayappan, P.},
  booktitle={Proc. 29th ACM Int. Conf. on Supercomputing (ICS)}, year={2015}}

@inproceedings{zhao2018bridging,
  title={Bridging the gap between deep learning and sparse matrix format selection},
  author={Zhao, Yue and Li, Jiajia and Liao, Chunhua and Shen, Xipeng},
  booktitle={Proc. 23rd ACM SIGPLAN Symp. on Principles and Practice of Parallel Programming (PPoPP)}, pages={94--108}, year={2018}}

@article{friedman2001greedy,
  title={Greedy function approximation: a gradient boosting machine},
  author={Friedman, Jerome H.},
  journal={Annals of Statistics}, volume={29}, number={5}, pages={1189--1232}, year={2001}}

@misc{openai,
  author       = {{OpenAI}},
  title        = {{GPT-5.6}: Frontier Intelligence That Scales with Your Ambition},
  year         = {2026},
  month        = jul,
  url          = {https://openai.com/index/gpt-5-6/},
  note         = {Accessed: 2026-07-31}
}

@inproceedings{reddit,
author = {Hamilton, William L. and Ying, Rex and Leskovec, Jure},
title = {Inductive representation learning on large graphs},
year = {2017},
isbn = {9781510860964},
publisher = {Curran Associates Inc.},
address = {Red Hook, NY, USA},
booktitle = {Proceedings of the 31st International Conference on Neural Information Processing Systems},
pages = {1025–1035},
numpages = {11},
location = {Long Beach, California, USA},
series = {NIPS'17}
}

@inproceedings{maxkgnn,
author = {Peng, Hongwu and Xie, Xi and Shivdikar, Kaustubh and Hasan, Md Amit and Zhao, Jiahui and Huang, Shaoyi and Khan, Omer and Kaeli, David and Ding, Caiwen},
title = {MaxK-GNN: Extremely Fast GPU Kernel Design for Accelerating Graph Neural Networks Training},
year = {2024},
isbn = {9798400703850},
publisher = {Association for Computing Machinery},
address = {New York, NY, USA},
url = {https://doi.org/10.1145/3620665.3640426},
doi = {10.1145/3620665.3640426},
booktitle = {Proceedings of the 29th ACM International Conference on Architectural Support for Programming Languages and Operating Systems, Volume 2},
pages = {683–698},
numpages = {16},
location = {La Jolla, CA, USA},
series = {ASPLOS '24}
}

@article{VAZQUEZ2010146,
title = {A matrix approach to tomographic reconstruction and its implementation on GPUs},
journal = {Journal of Structural Biology},
volume = {170},
number = {1},
pages = {146-151},
year = {2010},
issn = {1047-8477},
doi = {https://doi.org/10.1016/j.jsb.2010.01.021},
url = {https://www.sciencedirect.com/science/article/pii/S104784771000033X},
author = {F. Vázquez and E.M. Garzón and J.J. Fernández},
}

\clearpage
\appendix
\section{Compilation-Plan Design Space}
\label{app:strategy-space}

Table~\ref{tab:app-strategy} expands the compilation-plan formulation in
Section~\ref{sec:design}. It lists representative choices for the
representation $R=\langle L,I\rangle$, execution schedule
$S=\langle P,D\rangle$, and hardware mapping $\theta_H$ of each operator.
These choices are coupled: the representation constrains the parallel
decomposition and dataflow, which in turn affect hardware resource use.
Synthesis combines compatible choices into a complete plan. The entries
describe optimization mechanisms, not a fixed set of complete kernels.

\begin{table*}[t]
  \centering \footnotesize \setlength{\tabcolsep}{4pt}
  \begin{tabular}{@{}p{0.075\textwidth}p{0.205\textwidth}p{0.185\textwidth}p{0.215\textwidth}p{0.195\textwidth}@{}}
    \toprule
    \textbf{Op.} & \textbf{Layout and metadata $\langle L,I\rangle$}
    & \textbf{Parallel decomposition $P$} & \textbf{Dataflow $D$}
    & \textbf{Hardware mapping $\theta_H$} \\
    \midrule
    SpMV
      & CSR, COO, ELL, SELL-$C$-$\sigma$ with slice height $C$ and sort
        window $\sigma$, BSR, adaptive $16\times16$ tiles with per-tile
        format selection. Metadata: row pointers, slice offsets, block
        indices, row permutations, per-tile descriptors.
      & One thread, one warp, or one CTA per row; equal-nonzero chunks
        with merge-path boundaries; slice- or block-parallel; cyclic row
        assignment for balance.
      & Direct write to $y$; segmented reduction across chunk boundaries;
        atomic accumulation; read-only or texture-cached access to $x$;
        windowed staging of $x$ segments in shared memory.
      & Rows or nonzeros per CTA, slice height, block size, threads per
        block, active column-window width, unroll factor. \\
    \addlinespace
    SpMM
      & CSR, COO, BSR, Blocked-ELL, packed column windows, adaptive tiles.
        Metadata: row and block indices, window descriptors, Tensor-Core
        fragment maps.
      & Sparse rows, nonzeros, or blocks crossed with tiles of the dense
        dimension $K$; one warp per row window; two-level tiling over
        (rows, $K$).
      & Reuse of dense rows across sparse nonzeros; accumulation in
        registers or shared memory; $B$ staged in shared memory with $C$
        held in registers; CUDA-core or Tensor-Core (\texttt{mma})
        execution; double buffering of the sparse stream.
      & $K$ tile width, block shape, warps per CTA, shared-memory staging
        size, fragment layout, double-buffer depth. \\
    \addlinespace
    SpGEMM
      & CSR, row-reordered CSR, tiled inputs, materialized
        $A^{\mathsf{T}}$ for $AA^{\mathsf{T}}$. Metadata: row bins by
        upper-bound work, intermediate-product counts, tile pointers.
      & Rows grouped into bins by predicted output size; one warp or CTA
        per row; tile-parallel; separate symbolic and numeric phases.
      & Dense, hash, sorted-merge, or hybrid accumulation; on-chip
        accumulators sized per bin; global spill for oversized rows;
        output compaction by prefix sum.
      & Hash-table capacity, dense-stripe width, bin thresholds, threads
        per row, shared-memory allocation per CTA. \\
    \bottomrule
  \end{tabular}
  \caption{Representative compilation-plan choices by operator. Synthesis
  coordinates representation, parallel decomposition, dataflow, and
  hardware mapping rather than selecting them independently.}
  \label{tab:app-strategy}
\end{table*}

\section{Feature Importance}
\label{app:importance}

\begin{figure}[t]
\centering
\includegraphics[width=0.9\columnwidth]{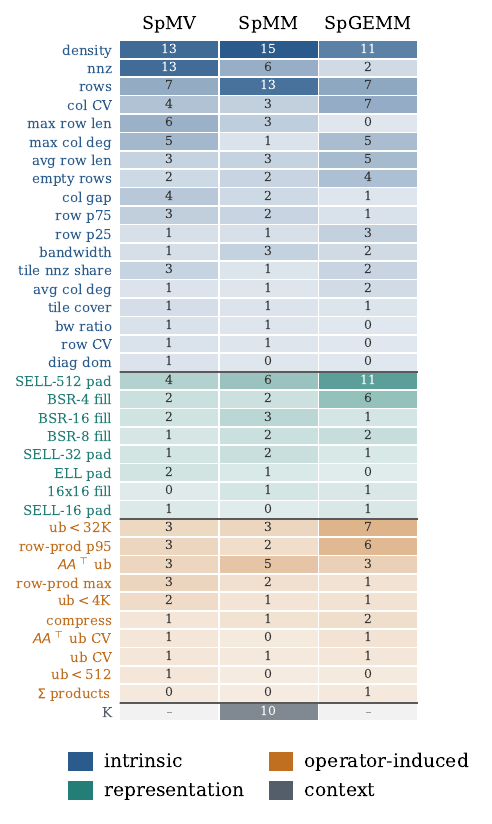}
\caption{Normalized model-native feature importance by operator, in
percent. Features are grouped by the analysis taxonomy and ordered
within each group. The dense width $K$ applies only to SpMM.}
\Description{A heatmap shows model-native importance for 36 structural
features and the SpMM dense width. Density and matrix size contribute
across operators, with additional operator-specific features.}
\label{fig:ebm-full}
\end{figure}

\paragraph{Model-native importance.}
The model in Equation~\ref{eq:additive-energy} exposes
each feature's score contribution,
$\varphi_{j,o,s}(x_j)=h_j(x_j)^{\mathsf{T}}w_{2,o,j,s}$.
Template rankings depend on differences between scores. We summarize a
feature's influence by the variation of its contribution across templates
at the mean feature vector $\bar{x}$:
\begin{equation}
  I_j(o)=\operatorname{std}_{s\in V_o}\;\varphi_{j,o,s}(\bar{x}_j),
  \label{eq:ebm-importance}
\end{equation}
The values are normalized to sum to one per operator.
Figure~\ref{fig:ebm-full} reports this measure for the $36$ structural
features and the SpMM width $K$. It complements the permutation
importance in Figure~\ref{fig:feat-importance}: one describes score
variation at a reference feature vector, while the other measures
sensitivity to feature permutation across inputs.

Density, nonzero count, and row count are influential across operators.
The dense width $K$ is prominent for SpMM, while intermediate-work
features matter for SpGEMM. About $30$ features account for $90\%$ of
the model-native importance, indicating that the scores draw on a broad
set of structural signals.

\section{Evaluation Matrices}
\label{app:matrices}

Table~\ref{tab:app-matrices} lists the $60$ SuiteSparse matrices used in
Section~\ref{sec:eval}, sorted by nonzero count. Mean and Max denote the
mean and maximum number of nonzeros per row. CV is the coefficient of
variation of row length, and density is $\operatorname{nnz}(X)/(mn)$.
The last column indicates whether the matrix is square.

\begin{table*}[tp]
  \centering
  \scriptsize
  \setlength{\tabcolsep}{3pt}
  \caption{The $60$ evaluation matrices, sorted by nonzero count.}
  \label{tab:app-matrices}
  \begin{tabular}{@{}l r r r r r r r c@{}}
  \toprule
  \textbf{Matrix} & \textbf{Rows} & \textbf{Cols} & \textbf{NNZ}
  & \textbf{Density} & \textbf{Mean} & \textbf{Max} & \textbf{CV}
  & \textbf{Sq.} \\
  \midrule
HB\_bcsstm19 & 817 & 817 & 817 & $1.2\!\times\!10^{-3}$ & 1.0 & 1 & 0.00 & yes \\
HB\_str\_0 & 363 & 363 & 2{,}454 & $1.9\!\times\!10^{-2}$ & 6.8 & 34 & 1.09 & yes \\
LPnetlib\_lp\_standata & 359 & 1{,}274 & 3{,}230 & $7.1\!\times\!10^{-3}$ & 9.0 & 745 & 4.41 & no \\
HB\_bp\_600 & 822 & 822 & 4{,}172 & $6.2\!\times\!10^{-3}$ & 5.1 & 302 & 2.37 & yes \\
Bai\_cdde2 & 961 & 961 & 4{,}681 & $5.1\!\times\!10^{-3}$ & 4.9 & 5 & 0.07 & yes \\
Sandia\_fpga\_dcop\_04 & 1{,}220 & 1{,}220 & 5{,}884 & $4.0\!\times\!10^{-3}$ & 4.8 & 36 & 0.83 & yes \\
Sandia\_fpga\_dcop\_11 & 1{,}220 & 1{,}220 & 5{,}892 & $4.0\!\times\!10^{-3}$ & 4.8 & 36 & 0.83 & yes \\
Sandia\_fpga\_dcop\_47 & 1{,}220 & 1{,}220 & 5{,}892 & $4.0\!\times\!10^{-3}$ & 4.8 & 36 & 0.83 & yes \\
JGD\_Homology\_n3c6-b10 & 675 & 2{,}511 & 7{,}425 & $4.4\!\times\!10^{-3}$ & 11.0 & 11 & 0.00 & no \\
Oberwolfach\_flowmeter0 & 9{,}669 & 9{,}669 & 9{,}669 & $1.0\!\times\!10^{-4}$ & 1.0 & 1 & 0.00 & yes \\
HB\_can\_838 & 838 & 838 & 10{,}010 & $1.4\!\times\!10^{-2}$ & 11.9 & 32 & 0.52 & yes \\
Sandia\_adder\_dcop\_44 & 1{,}813 & 1{,}813 & 11{,}245 & $3.4\!\times\!10^{-3}$ & 6.2 & 1{,}310 & 4.96 & yes \\
Sandia\_adder\_dcop\_54 & 1{,}813 & 1{,}813 & 11{,}246 & $3.4\!\times\!10^{-3}$ & 6.2 & 1{,}310 & 4.96 & yes \\
PARSEC\_Si2 & 769 & 769 & 17{,}801 & $3.0\!\times\!10^{-2}$ & 23.1 & 61 & 0.39 & yes \\
HB\_bcsstk09 & 1{,}083 & 1{,}083 & 18{,}437 & $1.6\!\times\!10^{-2}$ & 17.0 & 23 & 0.24 & yes \\
Gset\_G47 & 1{,}000 & 1{,}000 & 19{,}980 & $2.0\!\times\!10^{-2}$ & 20.0 & 34 & 0.22 & yes \\
JGD\_G5\_IG5-11 & 1{,}227 & 1{,}692 & 22{,}110 & $1.1\!\times\!10^{-2}$ & 18.0 & 111 & 0.83 & no \\
Newman\_hep-th & 8{,}361 & 8{,}361 & 31{,}502 & $4.5\!\times\!10^{-4}$ & 3.8 & 50 & 1.14 & yes \\
Gset\_G31 & 2{,}000 & 2{,}000 & 39{,}980 & $10.0\!\times\!10^{-3}$ & 20.0 & 40 & 0.23 & yes \\
JGD\_Franz\_Franz5 & 7{,}382 & 2{,}882 & 44{,}056 & $2.1\!\times\!10^{-3}$ & 6.0 & 6 & 0.04 & no \\
DIMACS10\_fe\_4elt2 & 11{,}143 & 11{,}143 & 65{,}636 & $5.3\!\times\!10^{-4}$ & 5.9 & 12 & 0.15 & yes \\
FIDAP\_ex14 & 3{,}251 & 3{,}251 & 66{,}775 & $6.3\!\times\!10^{-3}$ & 20.5 & 37 & 0.38 & yes \\
Rommes\_bips07\_3078 & 21{,}128 & 21{,}128 & 75{,}729 & $1.7\!\times\!10^{-4}$ & 3.6 & 27 & 0.76 & yes \\
GHS\_indef\_aug2d & 29{,}008 & 29{,}008 & 76{,}832 & $9.1\!\times\!10^{-5}$ & 2.6 & 4 & 0.36 & yes \\
Schenk\_IBMNA\_c-38 & 8{,}127 & 8{,}127 & 77{,}689 & $1.2\!\times\!10^{-3}$ & 9.6 & 889 & 1.73 & yes \\
Boeing\_msc04515 & 4{,}515 & 4{,}515 & 97{,}707 & $4.8\!\times\!10^{-3}$ & 21.6 & 27 & 0.23 & yes \\
FIDAP\_ex15 & 6{,}867 & 6{,}867 & 98{,}671 & $2.1\!\times\!10^{-3}$ & 14.4 & 18 & 0.30 & yes \\
Hollinger\_g7jac040 & 11{,}790 & 11{,}790 & 114{,}671 & $8.2\!\times\!10^{-4}$ & 9.7 & 120 & 1.50 & yes \\
Bindel\_ted\_B\_unscaled & 10{,}605 & 10{,}605 & 144{,}579 & $1.3\!\times\!10^{-3}$ & 13.6 & 49 & 0.89 & yes \\
ML\_Graph\_kmnist\_norm\_10NN & 10{,}000 & 10{,}000 & 156{,}932 & $1.6\!\times\!10^{-3}$ & 15.7 & 140 & 0.47 & yes \\
Schenk\_IBMNA\_c-50 & 22{,}401 & 22{,}401 & 193{,}625 & $3.9\!\times\!10^{-4}$ & 8.6 & 1{,}919 & 2.65 & yes \\
Rajat\_rajat22 & 39{,}899 & 39{,}899 & 197{,}264 & $1.2\!\times\!10^{-4}$ & 4.9 & 3{,}401 & 5.05 & yes \\
Toledo\_deltaX & 68{,}600 & 21{,}961 & 247{,}424 & $1.6\!\times\!10^{-4}$ & 3.6 & 83 & 1.63 & no \\
JGD\_Franz\_Franz11 & 47{,}104 & 30{,}144 & 329{,}728 & $2.3\!\times\!10^{-4}$ & 7.0 & 7 & 0.00 & no \\
JGD\_GL7d\_GL7d13 & 47{,}271 & 8{,}899 & 356{,}232 & $8.5\!\times\!10^{-4}$ & 7.5 & 14 & 0.34 & no \\
Quaglino\_viscoplastic2 & 32{,}769 & 32{,}769 & 381{,}326 & $3.6\!\times\!10^{-4}$ & 11.6 & 73 & 1.20 & yes \\
Andrianov\_net25 & 9{,}520 & 9{,}520 & 401{,}200 & $4.4\!\times\!10^{-3}$ & 42.1 & 139 & 0.77 & yes \\
GHS\_indef\_c-59 & 41{,}282 & 41{,}282 & 480{,}536 & $2.8\!\times\!10^{-4}$ & 11.6 & 3{,}090 & 2.04 & yes \\
Shen\_e40r0100 & 17{,}281 & 17{,}281 & 553{,}562 & $1.9\!\times\!10^{-3}$ & 32.0 & 62 & 0.49 & yes \\
JGD\_Homology\_ch7-8-b4 & 141{,}120 & 58{,}800 & 705{,}600 & $8.5\!\times\!10^{-5}$ & 5.0 & 5 & 0.00 & no \\
AMD\_G2\_circuit & 150{,}102 & 150{,}102 & 726{,}674 & $3.2\!\times\!10^{-5}$ & 4.8 & 6 & 0.13 & yes \\
Watson\_Baumann & 112{,}211 & 112{,}211 & 760{,}631 & $6.0\!\times\!10^{-5}$ & 6.8 & 7 & 0.06 & yes \\
QY\_case39 & 40{,}216 & 40{,}216 & 1{,}042{,}160 & $6.4\!\times\!10^{-4}$ & 25.9 & 20{,}024 & 12.20 & yes \\
Pereyra\_landmark & 71{,}952 & 2{,}704 & 1{,}151{,}232 & $5.9\!\times\!10^{-3}$ & 16.0 & 16 & 0.00 & no \\
Norris\_heart1 & 3{,}557 & 3{,}557 & 1{,}387{,}773 & $1.1\!\times\!10^{-1}$ & 390.2 & 1{,}120 & 0.32 & yes \\
DIMACS10\_ga2010 & 291{,}086 & 291{,}086 & 1{,}418{,}056 & $1.7\!\times\!10^{-5}$ & 4.9 & 85 & 0.61 & yes \\
CEMW\_vfem & 93{,}476 & 93{,}476 & 1{,}434{,}636 & $1.6\!\times\!10^{-4}$ & 15.3 & 28 & 0.20 & yes \\
Barabasi\_NotreDame\_actors & 392{,}400 & 127{,}823 & 1{,}470{,}404 & $2.9\!\times\!10^{-5}$ & 3.7 & 646 & 2.75 & no \\
Nemeth\_nemeth26 & 9{,}506 & 9{,}506 & 1{,}511{,}760 & $1.7\!\times\!10^{-2}$ & 159.0 & 193 & 0.16 & yes \\
JGD\_Homology\_ch7-9-b5 & 423{,}360 & 317{,}520 & 2{,}540{,}160 & $1.9\!\times\!10^{-5}$ & 6.0 & 6 & 0.00 & no \\
Andrianov\_net125 & 36{,}720 & 36{,}720 & 2{,}577{,}200 & $1.9\!\times\!10^{-3}$ & 70.2 & 231 & 0.95 & yes \\
TSOPF\_TSOPF\_FS\_b300\_c2 & 56{,}814 & 56{,}814 & 8{,}767{,}466 & $2.7\!\times\!10^{-3}$ & 154.3 & 27{,}742 & 6.23 & yes \\
Gupta\_gupta3 & 16{,}783 & 16{,}783 & 9{,}323{,}427 & $3.3\!\times\!10^{-2}$ & 555.5 & 14{,}672 & 2.22 & yes \\
LAW\_in-2004 & 1{,}382{,}908 & 1{,}382{,}908 & 16{,}917{,}053 & $8.8\!\times\!10^{-6}$ & 12.2 & 7{,}753 & 3.04 & yes \\
Janna\_CoupCons3D & 416{,}800 & 416{,}800 & 22{,}322{,}336 & $1.3\!\times\!10^{-4}$ & 53.6 & 76 & 0.16 & yes \\
Belcastro\_human\_gene1 & 22{,}283 & 22{,}283 & 24{,}669{,}643 & $5.0\!\times\!10^{-2}$ & 1107.1 & 7{,}939 & 1.27 & yes \\
DIMACS10\_asia\_osm & 11{,}950{,}757 & 11{,}950{,}757 & 25{,}423{,}206 & $1.8\!\times\!10^{-7}$ & 2.1 & 9 & 0.23 & yes \\
Belcastro\_mouse\_gene & 45{,}101 & 45{,}101 & 28{,}967{,}291 & $1.4\!\times\!10^{-2}$ & 642.3 & 8{,}032 & 1.33 & yes \\
Schenk\_AFE\_af\_shell10 & 1{,}508{,}065 & 1{,}508{,}065 & 52{,}672{,}325 & $2.3\!\times\!10^{-5}$ & 34.9 & 35 & 0.03 & yes \\
DIMACS10\_delaunay\_n24 & 16{,}777{,}216 & 16{,}777{,}216 & 100{,}663{,}202 & $3.6\!\times\!10^{-7}$ & 6.0 & 26 & 0.22 & yes \\
  \bottomrule
  \end{tabular}
\end{table*}

\section{Learned Cost Model Evaluation}
\label{app:selector}

This section supplements Section~\ref{sec:eval-selector} with the
evaluation protocol, metric definitions, and per-operator results for
the learned template-ranking model.

\paragraph{Protocol.}
The offline corpus contains $400$ SuiteSparse matrices with measured
template runtimes. Because the structural features of
a matrix are shared across its tasks, we report five-fold
cross-validation grouped by matrix and stratified by shape, averaged over
five seeds; a separate $100$-matrix test set is held out and never used
for tuning. We use four metrics per operator. \emph{Top-1} is the
fraction of inputs whose top-ranked template is the measured best.
\emph{Regret} is the geometric-mean slowdown of the top-ranked template
relative to the per-input oracle. \emph{Capture} is the fraction of the
oracle speedup realized by the ranking. \emph{Cliff} is the fraction of
inputs whose top choice is more than $2\times$ slower than the CSR
baseline. Regret and capture, rather than top-1, are the quantities that
reflect the performance cost of a ranking error, because selecting a
near-optimal template can preserve performance despite a top-1 error.

\paragraph{Baselines.}
We compare against a multilayer perceptron (MLP) ranker trained with
the same listwise objective, and gradient-boosted trees (GBDT), a strong
reference model for tabular data. All models use the same $36$
structural features and identical cross-validation folds. \emph{Energy
(listwise)} ablates our model by removing the margin term
($\lambda_o=0$), leaving the same architecture trained as a listwise
ranker.

\begin{table*}[t]
  \centering \footnotesize \setlength{\tabcolsep}{4pt}
  \begin{tabular}{@{}l ccc ccc ccc@{}}
    \toprule
    & \multicolumn{3}{c}{\textbf{SpMV}}
    & \multicolumn{3}{c}{\textbf{SpMM}}
    & \multicolumn{3}{c}{\textbf{SpGEMM}} \\
    \cmidrule(lr){2-4}\cmidrule(lr){5-7}\cmidrule(lr){8-10}
    \textbf{Model} & top1$\uparrow$ & regret$\downarrow$ & capt.$\uparrow$
                   & top1$\uparrow$ & regret$\downarrow$ & capt.$\uparrow$
                   & top1$\uparrow$ & regret$\downarrow$ & capt.$\uparrow$ \\
    \midrule
    MLP                 & 0.805 & 1.03 & 95\% & 0.715 & 1.17 & 76\% & 0.681 & 1.37 & 76\% \\
    GBDT$^\ddagger$     & 0.793 & 1.03 & 96\% & 0.716 & 1.09 & 87\% & 0.713 & 1.26 & 82\% \\
    \midrule
    Energy (listwise)   & 0.797 & 1.03 & 96\% & 0.716 & 1.14 & 80\% & 0.697 & 1.34 & 78\% \\
    \textbf{Energy (ours)} & 0.797 & 1.03 & 96\% & \textbf{0.719} & \textbf{1.13} & \textbf{81\%} & \textbf{0.702} & \textbf{1.32} & \textbf{79\%} \\
    \bottomrule
  \end{tabular}
  \caption{Template-ranking quality (5-fold grouped cross-validation,
  averaged over 5 seeds).
  $^\ddagger$ marks the black-box reference model. Bold marks where the
  energy model beats the MLP on SpMM and SpGEMM. Cliff rates are
  discussed in the text.}
  \label{tab:app-selector}
\end{table*}

\paragraph{Ranking quality.}
The energy model improves on the MLP in top-1 accuracy, regret, and
capture for SpMM and SpGEMM (Table~\ref{tab:app-selector}). Capture rises
from $76\%$ to $81\%$ for SpMM and from $76\%$ to $79\%$ for SpGEMM.
For SpMV, both models have regret $1.03$; the energy model improves
capture from $95\%$ to $96\%$, with slightly lower top-1 accuracy.

GBDT achieves lower regret and higher capture on SpMM and SpGEMM.
The additive model remains competitive in top-1 accuracy on SpMV and
SpMM while exposing feature-level score contributions to synthesis.
Cliff rates remain below $6\%$ for all models. The energy model's
rates are $0.0\%$ for SpMV and $0.5\%$ for SpMM, no higher than the
MLP's. Removing the margin term reduces top-1 accuracy and capture on
SpMM and SpGEMM. This supports weighting pairwise rankings by measured
speedup gaps.

\section{Case Studies}
\label{app:cases}

The following cases illustrate how input structure and target hardware
affect the generated implementation. Table~\ref{tab:app-cases} reports
two tasks per operator on the RTX PRO 6000 Blackwell GPU with CUDA
13.0. The SpMV and SpMM pairs contrast different row-length
distributions; the SpGEMM pair contrasts intermediate work and output
size. A final case compares the same SpMM task on both GPUs.
Speedups use cuSPARSE unless explicitly noted.

\begin{table*}[t]
  \centering \footnotesize \setlength{\tabcolsep}{4pt}
  \begin{tabular}{@{}l l r r r r r@{}}
    \toprule
    \textbf{Op.} & \textbf{Matrix} & \textbf{Mean} & \textbf{Max}
    & \textbf{CV} & \textbf{Kernel (ms)} & \textbf{Speedup} \\
    \midrule
    SpMV & JGD\_Homology\_ch7-9-b5     & 6.0  & 6      & 0.00 & 0.007 & $2.44\times$ \\
    SpMV & Barabasi\_NotreDame\_actors & 3.7  & 646    & 2.75 & 0.013 & $1.94\times$ \\
    \addlinespace
    SpMM & Schenk\_AFE\_af\_shell10 ($K{=}32$)  & 34.9 & 35 & 0.03 & 0.736 & $6.82\times$ \\
    SpMM & LAW\_in-2004 ($K{=}128$)            & 12.2 & 7{,}753 & 3.04 & 1.874 & $1.97\times$ \\
    \addlinespace
    SpGEMM & DIMACS10\_asia\_osm       & 2.1  & 9      & 0.23 & 3.771 & $6.45\times^\dagger$ \\
    SpGEMM & DIMACS10\_delaunay\_n24   & 6.0  & 26     & 0.22 & 22.009 & $1.19\times$ \\
    \bottomrule
  \end{tabular}
  \caption{Case-study tasks. Mean and Max are nonzeros per row, CV their
  coefficient of variation. Speedup is over cuSPARSE, except
  $^\dagger$asia\_osm, where cuSPARSE SpGEMM exits with an
  out-of-resources error and the reference is the HSMU-SpGEMM FP64
  build instead.}
  \label{tab:app-cases}
\end{table*}

\subsection{SpMV}

\paragraph{JGD\_Homology\_ch7-9-b5: uniform row lengths.}
This $423{,}360\times317{,}520$ matrix has exactly six nonzeros in every
row, with CV $0.00$. ELL therefore introduces no padding. The generated
plan uses column-major ELL and assigns one thread per row. Adjacent
threads access the same slot in adjacent rows through consecutive
addresses. The selector initially ranks SELL with slice height $32$
first. Synthesis instead produces an ELL implementation that avoids
slice metadata and achieves a $2.44\times$ speedup over cuSPARSE.

\paragraph{Barabasi\_NotreDame\_actors: skewed rows.}
The input contains $392{,}400$ rows and $1.47$M nonzeros. The mean row
length is $3.7$, the maximum is $646$, and CV is $2.75$.
Padding every row to the maximum would increase storage by roughly two
orders of magnitude. The generated implementation instead uses COO and
partitions the nonzero stream into equal-size chunks, with one nonzero
per thread. Warp-local segmented reductions combine contributions to
the same row, and atomic additions merge partial sums across warps.
This balances nonzero processing across warps and achieves
$1.94\times$. The two SpMV cases illustrate how row-length variation
changes both representation and parallel decomposition.

\subsection{SpMM}

\paragraph{Schenk\_AFE\_af\_shell10 at $K{=}32$: regular finite-element structure.}
This matrix has $1.5$M rows and $52.7$M nonzeros, with mean row length
$34.9$, maximum $35$, and CV $0.03$. ELL requires little padding.
The generated kernel assigns a four-thread group to each sparse row.
Each thread accumulates eight output values in registers, while shuffle
instructions share sparse entries within the group. The implementation
pipelines sparse loads and writes each output once the row is complete.
It achieves $6.82\times$. Although the selector ranks a Tensor-Core
block template first, measured refinement selects this CUDA-core
implementation. As in Section~\ref{sec:eval-cache}, template rankings
provide an initial prior rather than fixing the final implementation.

\paragraph{LAW\_in-2004 at $K{=}128$: a web graph.}
This matrix has $1.4$M rows and $16.9$M nonzeros, with mean row length
$12.2$, maximum $7{,}753$, and CV $3.04$. The generated implementation
separates short rows from long rows. A warp processes each short row
with register accumulation. Long rows are split into chunks of at most
$256$ nonzeros, each assigned to a CTA; atomic additions combine the
partial outputs. Both paths read the dense operand directly. This
hybrid mapping achieves $1.97\times$, with a compute time of
$1.874$~ms. Preprocessing takes $8.9$~s and is excluded from that
measurement. Its work lists depend on the sparse structure and can be
reused across subsequent calls with different dense operands.
Appendix~\ref{app:code} gives the generated source.

\subsection{SpGEMM}

\paragraph{DIMACS10\_asia\_osm: a sparse road network.}
This road network has $12.0$M rows, $2.1$ nonzeros per row on average,
and density $1.8\times10^{-7}$. The product $AA$ has $42.8$M nonzeros.
A full-width dense accumulator would require about $12$M entries per
row despite the sparse output. The generated plan instead bins rows by
their intermediate-product upper bounds. Rows with bounds of at most
$8$ or $36$ use fixed-capacity thread-local accumulators; larger rows
use a separate traversal-based path. The measured speedup is
$6.45\times$ against the HSMU-SpGEMM FP64 build. cuSPARSE exits with an
out-of-resources error on this task, so this comparison is separate
from the main cuSPARSE speedup aggregate.

\paragraph{DIMACS10\_delaunay\_n24: a large sparse output.}
This Delaunay triangulation has $16.8$M rows and $100.7$M nonzeros,
with six nonzeros per row on average. The generated plan bins rows by
their intermediate-product upper bounds and assigns a warp to each row.
The two bins use shared-memory hash tables with $64$ and $512$ entries,
respectively. A compaction pass produces the final output, which contains
$347$M nonzeros. The kernel achieves $1.19\times$. Binning reduces
accumulator overhead, but it does not remove the substantial traffic
required to materialize this output.

\subsection{The Same Task on Two GPUs}

The final case fixes the matrix and varies the GPU to illustrate
architecture-aware synthesis.

\texttt{Nemeth\_nemeth26} at $K{=}8$ is a $9{,}506$-row matrix with a
strongly diagonal structure. It has $159$ nonzeros per row on average
with CV $0.16$, a median row bandwidth of $185$ columns, and an average
gap of $1.2$ between consecutive column indices. Dense tiles account for
$65.2\%$ of the nonempty $16{\times}16$ tiles and contain $88.2\%$ of
the nonzeros. The structural profile and selector ranking are identical
on both GPUs, with a $32{\times}32$ Blocked-ELL Tensor-Core template
ranked first.

On the RTX PRO 6000, the selected implementation uses Tensor Cores and
achieves $4.57\times$ over cuSPARSE. On the H200, synthesis retains a
Tensor-Core candidate but also explores a SIMT block-row plan. The
latter wins at $8.33\times$.

The planning record illustrates how hardware capacities guide these
choices. It estimates a $9.28$~MiB working set for a $16{\times}16$ BSR
candidate, including dense operands and output, compared with
$12.15$~MiB for CSR. It compares these footprints with the H200's
$50$~MiB physical L2 capacity and considers an $8$~KiB shared-memory
tile for a windowed candidate. These estimates guide initial plans;
actual cache behavior and parallelism depend on the lowered code.
Target-GPU measurements then determine which implementation to retain.
The different winners show why an identical structural prior need not
lead to an identical implementation across GPUs.

\section{A Generated SpMM Implementation}
\label{app:code}

Listing~\ref{lst:app-spmm-web} gives the generated SpMM implementation
for LAW\_in-2004 at $K{=}128$, without manual edits. It follows the
harness's preprocessing, compute, and cleanup interface. Preprocessing
prepares the sparse representation and work lists. The compute entry
point evaluates the operator, and cleanup releases resources.
The reported $1.97\times$ speedup measures only the compute function,
as in the main evaluation.

Preprocessing separates rows shorter than $256$ nonzeros from chunks
of longer rows. The short-row kernel assigns one warp per row, with
four output accumulators per thread. The long-row kernel assigns a
$128$-thread CTA to each chunk and atomically combines partial outputs.
Both kernels read CSR and the dense operand directly. The generated
preprocessing also constructs a windowed representation, but the compute
function does not use it. The executed plan combines row-wise and
chunk-wise parallelism within one implementation.

\begin{lstlisting}[style=cuda,
  caption={Generated SpMM implementation for LAW\_in-2004 at $K{=}128$.},
  label={lst:app-spmm-web}]
#include <cuda_runtime.h>
#include <stdint.h>
#include <vector>
#include "sparse_formats.cuh"

struct SpmmContext {
    WindowedData windowed;
    int* d_light_rows;
    int light_count;
    int* d_chunk_rows;
    int* d_chunk_begin;
    int* d_chunk_end;
    int chunk_count;
    int num_rows;
    int nnz;
};

__global__ __launch_bounds__(256, 4)
void light_rows_spmm_kernel(
    const int* __restrict__ row_ids,
    int count,
    const int* __restrict__ row_ptr,
    const int* __restrict__ col_ind,
    const float* __restrict__ values,
    const float* __restrict__ B,
    float* __restrict__ C)
{
    const int tid = (int)threadIdx.x;
    const int warp = tid >> 5;
    const int lane = tid & 31;
    const int idx = (int)blockIdx.x * 8 + warp;

    if (idx >= count) return;

    const int row = row_ids[idx];
    const int begin = row_ptr[row];
    const int end = row_ptr[row + 1];

    float x0 = 0.0f;
    float x1 = 0.0f;
    float x2 = 0.0f;
    float x3 = 0.0f;

    for (int p = begin; p < end; ++p) {
        const int col = __ldg(col_ind + p);
        const float a = __ldg(values + p);
        const float* bp = B + ((size_t)col << 7) + lane;

        x0 = fmaf(a, __ldg(bp), x0);
        x1 = fmaf(a, __ldg(bp + 32), x1);
        x2 = fmaf(a, __ldg(bp + 64), x2);
        x3 = fmaf(a, __ldg(bp + 96), x3);
    }

    float* cp = C + ((size_t)row << 7) + lane;
    cp[0] = x0;
    cp[32] = x1;
    cp[64] = x2;
    cp[96] = x3;
}

__global__ __launch_bounds__(128, 6)
void compact_dense_chunks_spmm_kernel(
    const int* __restrict__ chunk_rows,
    const int* __restrict__ chunk_begin,
    const int* __restrict__ chunk_end,
    int chunk_count,
    const int* __restrict__ col_ind,
    const float* __restrict__ values,
    const float* __restrict__ B,
    float* __restrict__ C)
{
    const int task = (int)blockIdx.x;
    if (task >= chunk_count) return;

    const int lane = (int)threadIdx.x;
    const int row = chunk_rows[task];
    const int begin = chunk_begin[task];
    const int end = chunk_end[task];

    float s = 0.0f;

    for (int p = begin; p < end; ++p) {
        const int col = __ldg(col_ind + p);
        const float a = __ldg(values + p);
        s = fmaf(a, __ldg(B + ((size_t)col << 7) + lane), s);
    }

    atomicAdd(C + ((size_t)row << 7) + lane, s);
}

void* spmm_preprocess(
    const int* d_row_ptr,
    const int* d_col_ind,
    const float* d_values,
    int num_rows,
    int num_cols,
    int nnz,
    int K)
{
    (void)K;

    SpmmContext* ctx = new SpmmContext;
    ctx->d_light_rows = nullptr;
    ctx->d_chunk_rows = nullptr;
    ctx->d_chunk_begin = nullptr;
    ctx->d_chunk_end = nullptr;
    ctx->light_count = 0;
    ctx->chunk_count = 0;
    ctx->num_rows = num_rows;
    ctx->nnz = nnz;

    ctx->windowed = csr_to_windowed(
        d_row_ptr, d_col_ind, d_values,
        num_rows, num_cols, nnz, 256, 4);

    if (num_rows <= 0)
        return ctx;

    std::vector<int> h_row_ptr((size_t)num_rows + 1);
    cudaMemcpy(h_row_ptr.data(), d_row_ptr,
               ((size_t)num_rows + 1) * sizeof(int),
               cudaMemcpyDeviceToHost);

    std::vector<int> h_light_rows;
    std::vector<int> h_chunk_rows;
    std::vector<int> h_chunk_begin;
    std::vector<int> h_chunk_end;

    h_light_rows.reserve(num_rows);

    const int chunk_nnz = 256;
    for (int r = 0; r < num_rows; ++r) {
        const int begin = h_row_ptr[r];
        const int end = h_row_ptr[r + 1];
        const int n = end - begin;

        if (n < chunk_nnz) {
            h_light_rows.push_back(r);
        } else {
            for (int p = begin; p < end; p += chunk_nnz) {
                int q = p + chunk_nnz;
                if (q > end) q = end;
                h_chunk_rows.push_back(r);
                h_chunk_begin.push_back(p);
                h_chunk_end.push_back(q);
            }
        }
    }

    ctx->light_count = (int)h_light_rows.size();
    ctx->chunk_count = (int)h_chunk_rows.size();

    if (ctx->light_count > 0) {
        cudaMalloc(&ctx->d_light_rows,
                   (size_t)ctx->light_count * sizeof(int));
        cudaMemcpy(ctx->d_light_rows, h_light_rows.data(),
                   (size_t)ctx->light_count * sizeof(int),
                   cudaMemcpyHostToDevice);
    }

    if (ctx->chunk_count > 0) {
        const size_t bytes = (size_t)ctx->chunk_count * sizeof(int);
        cudaMalloc(&ctx->d_chunk_rows, bytes);
        cudaMalloc(&ctx->d_chunk_begin, bytes);
        cudaMalloc(&ctx->d_chunk_end, bytes);

        cudaMemcpy(ctx->d_chunk_rows, h_chunk_rows.data(),
                   bytes, cudaMemcpyHostToDevice);
        cudaMemcpy(ctx->d_chunk_begin, h_chunk_begin.data(),
                   bytes, cudaMemcpyHostToDevice);
        cudaMemcpy(ctx->d_chunk_end, h_chunk_end.data(),
                   bytes, cudaMemcpyHostToDevice);
    }

    return ctx;
}

void spmm_compute(
    void* context,
    const int* d_row_ptr,
    const int* d_col_ind,
    const float* d_values,
    const float* d_B,
    float* d_C,
    int num_rows,
    int num_cols,
    int nnz,
    int K)
{
    (void)num_cols;
    (void)nnz;
    (void)K;

    SpmmContext* ctx = static_cast<SpmmContext*>(context);
    if (!ctx || num_rows <= 0)
        return;

    if (ctx->nnz == 0) {
        cudaMemsetAsync(d_C, 0, (size_t)num_rows * 128 * sizeof(float), 0);
        return;
    }

    if (ctx->chunk_count > 0) {
        cudaMemsetAsync(d_C, 0, (size_t)num_rows * 128 * sizeof(float), 0);

        if (ctx->light_count > 0) {
            const int blocks = (ctx->light_count + 7) >> 3;
            light_rows_spmm_kernel<<<blocks, 256>>>(
                ctx->d_light_rows,
                ctx->light_count,
                d_row_ptr,
                d_col_ind,
                d_values,
                d_B,
                d_C);
        }

        compact_dense_chunks_spmm_kernel<<<ctx->chunk_count, 128>>>(
            ctx->d_chunk_rows,
            ctx->d_chunk_begin,
            ctx->d_chunk_end,
            ctx->chunk_count,
            d_col_ind,
            d_values,
            d_B,
            d_C);
    } else if (ctx->light_count > 0) {
        const int blocks = (ctx->light_count + 7) >> 3;
        light_rows_spmm_kernel<<<blocks, 256>>>(
            ctx->d_light_rows,
            ctx->light_count,
            d_row_ptr,
            d_col_ind,
            d_values,
            d_B,
            d_C);
    }
}

void spmm_cleanup(void* context)
{
    SpmmContext* ctx = static_cast<SpmmContext*>(context);
    if (!ctx)
        return;

    if (ctx->d_light_rows)
        cudaFree(ctx->d_light_rows);
    if (ctx->d_chunk_rows)
        cudaFree(ctx->d_chunk_rows);
    if (ctx->d_chunk_begin)
        cudaFree(ctx->d_chunk_begin);
    if (ctx->d_chunk_end)
        cudaFree(ctx->d_chunk_end);

    free_windowed(ctx->windowed);
    delete ctx;
}
\end{lstlisting}

\end{document}